\documentclass[aps,nofootinbib,superscriptaddress, showpacs,preprintnumbers, nofootinbibt,twocolumn]{revtex4-2}
\usepackage{booktabs}
\usepackage[utf8]{inputenc}
\usepackage[T1]{fontenc}
\usepackage{amsmath}
\usepackage{amssymb}
\usepackage{graphicx}
\usepackage{caption}
\usepackage{subcaption}
\usepackage[most]{tcolorbox}
\usepackage{mathtools}
\usepackage{epsfig}
\usepackage{multirow}
\usepackage{eurosym}
\usepackage{dcolumn}
\usepackage{bm}
\usepackage{enumerate}
\usepackage{float}
\usepackage{epstopdf}
\usepackage{amsmath}
\usepackage{bm}
\usepackage{amsfonts}
\usepackage{amssymb}
\usepackage{graphicx}
\usepackage{alphalph,mathtools}
\usepackage{etoolbox}
\usepackage{color}
\usepackage{booktabs}
\usepackage{footnote}
\usepackage{makecell,tabularx}
\usepackage[leftcaption]{sidecap}
\usepackage{amssymb}
\usepackage{graphicx}
\usepackage{booktabs,dcolumn,caption}
\usepackage{rotating}
\usepackage{amsthm}
\usepackage{tikz}
\usepackage{caption}
\usepackage{subcaption}
\usepackage[most]{tcolorbox}

\usepackage{hyperref}
\hypersetup{colorlinks,citecolor=green}
\hypersetup{colorlinks=true,linkcolor=cyan,filecolor=green, urlcolor=magenta}

\def\be{\begin{eqnarray}}
	\def\ee{\end{eqnarray}}
\def\bea{\begin{eqnarray}}
	\def\eea{\end{eqnarray}}

\begin{document}
\title{Conservative wormholes in generalized $\kappa(\mathcal{R},\mathcal{T})$-
function}		
\author{Ksh. Newton Singh}
\email[Email:]{ntnphy@gmail.com}
\affiliation{Department of Physics, National Defence Academy, Khadakwasla, Pune 411023, India }

\author{G. R. P.  Teruel}
\email{gines.landau@gmail.com}
\affiliation{Departamento de Matemáticas, IES Carrús, Elche 03205, Alicante, Spain.}

\author{S. K. Maurya}
\email[Email:]{sunil@unizwa.edu.om}
\affiliation{Department of Mathematical and Physical Sciences,
College of Arts and Sciences, University of Nizwa, P.O. Box 33, Nizwa 616, Sultanate of Oman}

\author{Tanmoy Chowdhury}
\email{tanmoych.ju@gmail.com}
\affiliation{Department of Mathematics, Jadavpur University, Kolkata 700032, West Bengal, India.}

\author{Farook Rahaman}
\email{rahaman@associates.iucaa.in}
\affiliation{Department of Mathematics, Jadavpur University, Kolkata 700032, West Bengal, India.}

\begin{abstract}
We present an exhaustive study of wormhole configurations in $\kappa(\mathcal{R},\mathcal{T})$ gravity with linear and non-linear functions. The model assumed Morris-Thorne spacetime where the redshift and shape functions linked with the matter contain and geometry of the spacetime through non-covariant conservation equation of the stress-energy tensor. The first solution was explored assuming a constant redshift function that leads to a wormhole (WH) which is asymptotically non-flat. The remaining solutions were explored in two cases. Firstly, assuming a linear equation of state $p(r)=\omega \rho(r)$ along with different forms of $\kappa(\mathcal{R},\mathcal{T})-$function. This proved enough to derive a shape function of the form $b(r)=r_{0}\left(\frac{r_{0}}{r}\right)^{1/\omega}$. Secondly, by assuming specific choices of the shape function consistent with the wormhole configuration requirements. All the solutions fulfill flare-out condition, asymptotically flat and supported by phantom energy. Further, the embedding surface and its revolution has been generated using numerical method to see how the length of the throat is affected of the coupling parameters through $\kappa(\mathcal{R},\mathcal{T})$ function. At the end, we have also calculated the average null energy condition, which is satisfied by all the WH models signifying minimum exotic matter is required to open the WH throats.\\

\textbf{Keywords:} Exact solutions; Wormholes; $\kappa(\mathcal{R},\mathcal{T})$ modified gravity theory; Stellar structure

\end{abstract}
		
	\maketitle
\tableofcontents

\section{INTRODUCTION}
The theory of General Relativity (GR) as well as other extended theories of gravity allow for a spacetime that contains a complex structure, like a wormhole. A wormhole is a tunnel-like structure that links two faraway or distinct spaces. Wormholes and black holes are interesting astrophysical objects in GR. The presence of black holes has previously been confirmed~\cite{Wr1,Wr2,Wr3}. However, the presence of wormholes in the cosmos is a subject of ongoing research. Einstein and Rosen~\cite{Wr4} came up with the first wormhole solution which is referred to as the Einstein-Rosen bridge. Due to a non-traversable structure, this wormhole was considered to be only a mathematical model. After many years, Ellis~\cite{Wr5} discovered a novel wormhole solution for a spherically symmetric configuration of Einstein's equations, including a massless scalar field with ghost properties. Morris and Thorne~\cite{Wr6} show that such Ellis wormholes are traversable which allows for instantaneous travel across space and the possibility of time travel. Such wormhole models do not possess a singularity or a horizon, and the tidal force is sufficiently small for humans to survive it.  Furthermore,  Morris and Thorne~\cite{Wr6} also confirmed that the wormhole solution in GR should violate the null energy conditions that require the exotic matter.  This exotic matter, which violates energy conditions, has physical characteristics that would contradict established principles of physics, such as a particle exhibiting a negative mass.  Khatsymovsky ~\cite{Wr7} did an extensive study on the existence of wormholes. 
Several researchers have examined the stability of traversable wormholes. In this connection, Shinkai and Hayward~\cite{Wr8} showed via numerical simulations that Ellis wormholes exhibit instability. Since the inception of the traversable wormhole model, the feasibility of constructing wormholes with ordinary matter has intrigued researchers. Recent studies suggest that in modified gravity theories, it might be possible to create wormholes composed of ordinary matter that adhere to all energy conditions \cite{epl_paper}. However, while using modified gravity to form wormholes, the matter may be ordinary, the effective geometric matter, the source of modified gravity, can still violate the usual null energy condition. Various studies have identified wormholes that do not require exotic matter \cite{fukutaka1989,hochberg1990,ghoroku1992,furey2005,bronnikov2010,kanti2011,kanti2012,harko2013,moraes2018,godani2021,sengupta2022}.

It is worthwhile looking into the background of some of these modified gravity theories. In recent years, significant progress has been made in modified theories of gravity, with researchers exploring various extensions of GR. One such extension is $f(\mathcal{R})$ gravity, where the standard Einstein-Hilbert action is modified by replacing the Ricci scalar with a function of the scalar curvature \cite{Buchdahl11,Starobinsky11,Felice11}.  This modification introduces changes to the equations that govern the gravitational field, which may have implications for the behavior of gravity at various scales. The theory of gravity, denoted as $f(\mathcal{R})$, has been suggested as a potential explanation for the observed acceleration of the Universe's expansion. It offers alternative explanations for phenomena such as stellar dynamics, galaxy rotation curves, and galaxy morphology. Additionally, it provides valuable insights into the limitations of primordial inflation \cite{Capozziello/2012}. Lobo et al. \cite{Oliveira3} conducted a study whereby they constructed traversable wormhole geometries within the framework of $f(\mathcal{R})$ gravity. This was achieved by making assumptions about certain shape functions and equations of state (EoS). Banerjee et al. \cite{Banerjee} conducted a study on non-commutative wormholes under Lorentzian distributions within the context of $f(\mathcal{R})$ gravity. Furthermore, the researchers Shamir et al. \cite{R3} conducted a study on traversable wormhole solutions within the extension of $f(\mathcal{R})$ gravity, referred to as $f(\mathcal{R}, G)$ gravity. Similarly, Banerjee et al. \cite{Baner,ban} examined wormhole solutions in $f(\mathcal{R}, \mathcal{T})$ gravity that meet the null energy condition under isotropic pressure.
 
The area of modified gravity theories has seen a significant development known as $f(Q)$ gravity. This concept was first proposed by Jimenez et al. in 2018 as an extension of $f(\mathcal{R})$ gravity~\cite{Jimenez}. The gravitational field in $f(Q)$ gravity is determined only by the non-metricity scalar $Q$. The aforementioned hypothesis has effectively examined a range of observational datasets \cite{Soudi, Banos, Salzano, Koivisto11}. Additionally, there is evidence indicating that the force of gravity, denoted as $f(Q)$, may provide a challenge to the conventional $\Lambda$CDM model \cite{Anagnostopoulos}. The theory of gravity, denoted as $f(Q)$, has been successfully used in the investigation of astrophysical entities, including black holes \cite{Fell} and spherically symmetric configurations \cite{Zhai}. A study of wormhole geometries in the framework of $f(Q)$ gravity was done by Hassan et al.~\cite{Hassan11}, using a variety of equations of state, such as linear and non-linear models. The researchers were able to find exact solutions for the linear model and confirmed that, in this specific case, a small amount of exotic matter is needed for a traversable wormhole via the use of volume integral analysis. Mustafa et al. \cite{Mustafa} obtained wormhole solutions by applying the Karmarkar condition to the $f(Q)$ gravity function, indicating the potential for creating wormholes that meet energy requirements. Recent investigations into wormhole solutions for both distributions in $f(Q)$ gravity and other modified theories of gravity \cite{Sokoliuk, Hassan1, Song11, Rahman11, Feng11, Baruah11, Sharif11, Rahaman22, Jamil22, Rani22,Moraes,Tripathy:2023ped,Sahoo:2023dus,Agrawal:2022atn,Agrawal:2022atn,Mishra:2021ato,Javed:2024gog,Ashraf:2022ojr,Javed:2021obd}.

A recent addition to the field is $f(Q,\mathcal{T})$ gravity, which introduces a matter-geometry coupling where the Lagrangian is a function of both the non-metricity scalar $Q$ and the trace of the energy-momentum tensor $T$ \cite{Y.Xu}. Despite being a developing theory, $f(Q,\mathcal{T})$ gravity has shown promise in various cosmological contexts. Studies have explored its implications for late-time accelerated expansion with observational constraints \cite{Arora111}, cosmological inflation \cite{Shiravand}, baryogenesis \cite{Bhattacharjee}, cosmological perturbations \cite{Najera}, and the reconstruction of the $f(Q,\mathcal{T})$ Lagrangian \cite{Gadbail}. Nevertheless, the investigation of $f(Q,\mathcal{T})$ gravity in the field of astrophysics has been somewhat limited. The authors investigated the static spherically symmetric wormhole solutions in $f(Q,\mathcal{T})$ gravity for both linear and non-linear models, considering different equations of state. They discovered that exact solutions were attainable for the linear model, but posed challenges for the non-linear model~\cite{Tayde33}. Moreover, research has examined wormhole solutions in higher dimensions inside Gaussian distributions, observing their prevalence mainly in four and five dimensions \cite{Islam}. Additionally, strong and weak gravitational lensing has proven to be a potent tool for analyzing gravitational fields around various astrophysical objects, including black holes and wormholes. Several studies have utilized gravitational lensing to investigate wormholes in theoretical physics and astrophysics \cite{A1,A2,A3,A4,A5,A6,A7,A8,A9,A10}. Singh et al. \cite{kns1} have discussed wormhole solutions in $f(\mathbb{T})-$gravity under conformal symmetry method. In an interesting paper, shadow of a rotating wormhole solution was presented by Rahaman et al. \cite{raa}. 

The common footing of these modified theories of gravity is that they are Lagrangian theories, in the sense that all these proposals are formulated by different generalizations of the Einstein-Hilbert action. A natural question arises, whether it could be possible to take a different approach and work with a manifest non-Lagrangian theory. Non-Lagrangian theories are being investigated in the last years in several sectors of theoretical physics \cite{Gukov,Gadde,Tar}. For example, it is becoming appreciated that quantum field theories (QFT) without a traditional Lagrangian description deserve attention, not only because they seem to populate much of the QFT landscape, but also since they seem to offer opportunities in the search of new types of 4-manifold invariants. A non-Lagrangian modified gravitational theory dubbed as $\kappa(\mathcal{R},\mathcal{T})$-gravity was invented in 2018 \cite{Ter}. The motivation to openly depart from the traditional modified gravity program and openly consider a non-Lagrangian theory lies in the following observation: Neither of the two great classical field theories, Maxwell Electrodynamics and GR, were discovered by means of the variational method. Regarding the GR case, the field equations were first obtained following two guiding principles: the principle of general covariance and the equivalence principle. The action principle, i.e, the Einstein-Hilbert action, was discovered and incorporated to the theory in a final stage, when the correct field equation had already been derived. In addition, the current overpopulation of Lagrangian theories could be diverting physicists attention from other viable alternatives, and even hindering the discovery of new physical principles. For a detailed discussion on this and other aspects of the theory, the reader is referred to \cite{Ter}.

Since its publication in 2018, several works have been devoted to explore the implications of some specific models of $\kappa(\mathcal{R},\mathcal{T})$-gravity. In particular, the model $\kappa(\mathcal{T})=8\pi-\lambda\mathcal{T}$ has received the main attention. Pradhan and Ahmed \cite{ahm22} studied its cosmological implications, showing that it can account for the current scenario of an accelerating universe, including a very small value of the cosmological constant, which is in line with experimental data. Pradhan et al. \cite{arc22} further investigated a more complete cosmological scenario, while Dixit et al.\cite{arc23} explored the thermodynamic properties of the expansion of the cosmos in the context of this theory. Sarkar et al. \cite{sark19}, introduced the first study of wormhole in $\kappa(\mathcal{R},\mathcal{T})-$ gravity, while Teruel et al. \cite{GRP} published the first solutions modeling compact stars in $\kappa(\mathcal{R},\mathcal{T})$-gravity for isotropic coordinates. Taṣer and Dogru \cite{Tas23} investigated a more general, Krori-Barua compact star model within the context of this specific $\kappa(\mathcal{T})$ theory. All these articles are restricted to one particular selection of the $\kappa(\mathcal{T})$ functional, the case $\kappa(\mathcal{R},\mathcal{T}) \equiv \kappa(\mathcal{T})=8\pi-\lambda\mathcal{T}$, which corresponds to a matter-matter coupling, which generates an additional contribution in the field equation quadratic in the trace of the stress-energy tensor. 

Also very recently, Teruel et al. \cite{Ter2} investigated gravastar configurations in $\kappa(\mathcal{R},\mathcal{T})$-gravity, for a variety of particular models. They showed that the interior solutions are regular everywhere regardless of the specific form of the $\kappa(\mathcal{R},\mathcal{T})$ functional chosen.\\ 
The aim of this work is to investigate wormhole structures for a variety of linear and non-linear $\kappa(\mathcal{R},\mathcal{T})$ functionals. Therefore, the scope of this investigation is much more general than the wormhole investigation carried out by Sarkar et al. \cite{sark19}, which was only restricted to the model $\kappa(\mathcal{T})=8\pi-\lambda\mathcal{T}$. Another relevant aspect of the theory that is included in the present work and was absent in the aforementioned is the following: the non-conservation equation that satisfies the stress-energy tensor is taken into account in the derivation of the solutions, as we will discuss in the following sections.

\section{BACKGROUND OF $\kappa(\mathcal{R},\mathcal{T})-$ GRAVITY }\label{sec2}
The theoretical framework of the system depends upon the subsequent field equations, 
\begin{eqnarray}\label{FieldEquations}
R_{\mu\nu}-\frac{1}{2}\mathcal{R}~g_{\mu\nu}-\Lambda g_{\mu\nu}=\kappa(\mathcal{R},\mathcal{T})\,T_{\mu\nu},
\end{eqnarray}
where the cosmological constant is denoted by $\Lambda$, the space-time metric is $g_{\mu\nu}$, the Ricci tensor is $R_{\mu\nu}$, and the material content's stress-energy tensor is $T_{\mu\nu}$, while $\kappa(\mathcal{R},\mathcal{T})$ is a mathematical expression that represents a generalized Einstein's gravitational constant. We express it as a function of the traces $\mathcal{T} \equiv g_{\mu\nu}T^{\mu\nu}$, and $\mathcal{R} \equiv g_{\mu\nu}R^{\mu\nu}$. The inclusion of the functional $\kappa(\mathcal{R},\mathcal{T})$ allows for the examination of the possible existence of a running gravitational constant, but not within the framework of the variational technique. Brans and Dicke \cite{Brans1,Brans2} conducted a study on the possible inclusion of a variable, namely Einstein's gravitational constant, in the action. Their research yielded a theory that deviates significantly from the field equations (\ref{FieldEquations}).

The field equations (\ref{FieldEquations}) indicate that  $T_{\mu\nu}$ is not-covariantly conserved. The disappearance of the divergence on the left-hand side of Einstein's field equation leads to the following outcome, 
\begin{eqnarray}\label{nonconservation1}
\nabla{^\nu}\Big[\kappa(\mathcal{R},\mathcal{T})\,T_{\mu\nu}\Big]=0.
\end{eqnarray}
The non-conservation of the term $T_{\mu\nu}$ may be mathematically represented by the equation $\kappa(\mathcal{R},\mathcal{T})\neq 0$, as

\begin{eqnarray}\label{nonconservation2}
\nabla^{\nu}T_{\mu\nu}=-\frac{\nabla^{\nu} \kappa(\mathcal{R},\mathcal{T})}{\kappa(\mathcal{R},\mathcal{T})}~T_{\mu\nu}~~, ~~~~~\forall~~~~\kappa(\mathcal{R},\mathcal{T})\neq0. 
\end{eqnarray} 
When the value of $\kappa(\mathcal{R},\mathcal{T})=0$, the magnitude of the right-hand side of Einstein's field equations becomes zero. This intriguing phenomenon is likely to occur within some models of the early cosmos, given sufficiently high density. It would suggest an exponential expansion propelled by a cosmological constant. Naturally, the theory may be transformed into a more cautious structure by establishing a novel stress-energy tensor that is more efficient.
\begin{eqnarray}
S_{\mu\nu}=\kappa(\mathcal{R},\mathcal{T})\, T_{\mu\nu}.
\end{eqnarray}
The structure of the field equations is then obtained as, 
\begin{eqnarray}\label{nonconservation}
R_{\mu\nu}-\frac{1}{2}\mathcal{R}~g_{\mu\nu}-\Lambda\, g_{\mu\nu}=S_{\mu\nu},
\end{eqnarray}
Furthermore, the Bianchi identities suggest that Eq.(\ref{nonconservation}) is
\begin{eqnarray}
\nabla^{\nu}S_{\mu\nu}=0.
\end{eqnarray}
Subsequently, this theory exhibits a comparable formal framework to GR, but with a little alteration in the material composition. Prominent non-conservative gravitational theories include Rastall's gravity theory \cite{Rastall} and the Lagrangian theory proposed by Harko et al.\cite{Harko}.

It is worth mentioning that several scientists \cite{Lind, Viss} have raised criticism on some non-conservative gravity theories, such as the Rastall gravitational theory. The main point is the following: This type of theories has an equivalent structure that Einstein's gravity, with an identical geometrical sector, and a matter sector that incorporates a conserved effective stress-energy tensor $T^{eff}_{\mu\nu}$, that can be build purely from the matter sources. Furthermore, these theories cannot be considered truly alternative theories of gravity if the effective stress-energy tensor of the matter sources is independent of the space-time curvature, as is the case with Rastall gravity.  Consequently, the issue of the non-conservation of the stress-energy tensor falls within the scope of special relativity.
However, the stress-energy tensor of $\kappa(\mathcal{R},\mathcal{T})$ gravity is not primarily dictated by matter sources alone, since it may also be influenced by the space-time curvature via the trace $\mathcal{R}$. Hence, it may be inferred that the theory of $\kappa(\mathcal{R},\mathcal{T})$ can be seen as a modification of the GR framework, rather than a simple reinterpretation of its matter sector.

The following are some pertinent aspects of the theory: 
\begin{itemize}
\item The procedure of generalization alters the right-hand side of Einstein's field equations. Therefore, only the material content sector is generalized. Consequently, the equations will exhibit second-order behaviour in the metric coefficients, and the theory will not suffer from the usual instability that afflicted many of the higher-order gravitational theories.
\item The theory may be simplified to GR in the absence of matter sources, since the pure geometrical sector is equivalent to GR. 
\item  The dependence on $\mathcal {T}$ implies that, in specific instances of the $\kappa(\mathcal{T})=8\pi-\lambda\mathcal{T}$ type, or for a wider functional that directly links the $\mathcal{R}$ and $\mathcal{T}$ traces, the theory would yield identical predictions to GR when associated to standard (traceless) electromagnetic fields. 
 In this situation, only significant departures are expected for non-linear electrodynamics, where $\mathcal {T}\neq0$.
\end{itemize}

\section{FIELD EQUATIONS IN  $\kappa(\mathcal{R},\mathcal{T})-$ GRAVITY}\label{sec3}

The geometry of the wormhole is given by Morris-Thorne metric given by 
\begin{small}
\begin{eqnarray}
ds^{2}=-e^{2f}dt^{2}+\left[1-\frac{b}{r}\right]^{-1}dr^{2} +r^{2}(d\theta^{2}+\sin^{2}\theta \,d\phi^{2}) . 
\end{eqnarray} 
\end{small}
Assuming $\Lambda=0$, the field equations for any $\kappa(\mathcal{R},\mathcal{T})$ function are
 \begin{eqnarray}\label{array}
\rho(r)~\kappa(\mathcal{R},\mathcal{T}) &=& \frac{b'(r)}{r^2},  \\
p(r)~\kappa(\mathcal{R},\mathcal{T}) &=& \frac{2f'(r)}{r}\left(1-\frac{b(r)}{r}\right)-\frac{b(r)}{r^3}.
\end{eqnarray}
Further, the stress-energy tensor satisfied a non-covariant conservation equation acquires the form
\begin{footnotesize}
\begin{eqnarray}\label{nonconservation3}
p~\frac{d \kappa(\mathcal{R},\mathcal{T})}{dr}+\kappa(\mathcal{R},\mathcal{T})\left[\frac{d f(r)}{dr}\{\rho+p\}+\frac{d p}{dr}\right]=0,
\end{eqnarray}
\end{footnotesize}
where $f(r)$ and $b(r)$ have their usual meanings. For a physical WH, the following properties should be fulfilled:
\begin{itemize}
\item For an event horizon free, the redshift function must be non-singular and non-zero everywhere.
\item Flare-out condition, $0<b'(r_{0})<1$ at the throat $r=r_{0}$ must hold.
\item $b(r)<r$ for $r>r_{0}$.
\item The WH should be asymptotically flat i.e. $b(r)/r\to 0$ as $r\to\infty$.
\end{itemize}
Due to 4 unknowns $\rho(r)$, $p(r)$, $b(r)$, $f(r)$ and 3 equations, we have therefore freedom to arbitrarily choose one of them. We remark that Eq. (\ref{nonconservation3}) should be included in the analysis. With this equation, the degrees of freedom are reduced to one. In ref. \cite{Sarkar} this equation was omitted, and therefore the solutions obtained therein satisfy field equations, but they do not necessarily satisfy Eq. (\ref{nonconservation3}). We claim that consistent solutions representing wormhole configurations in $\kappa(\mathcal{R},\mathcal{T})-$ gravity should satisfy these three equations simultaneously.

\section{WORMHOLE CONFIGURATIONS IN $\kappa(\mathcal{R},\mathcal{T})-$ THEORY}

\subsection{Constant Redshift Function}
The constant redshift function condition is extensively employed in the literature because its simplicity. Hence, setting $f(r)=\text{constant}$ and $f^{\prime}=0$, we have that Eq. (\ref{nonconservation3}) becomes
\begin{eqnarray}
p(r)~\frac{d \kappa(\mathcal{R},\mathcal{T})}{dr}+\kappa(\mathcal{R},\mathcal{T})\frac{d p(r)}{dr}=0,
\end{eqnarray}
which can be recast in the equivalent, more compact form
\begin{eqnarray}
\frac{d}{dr}\Big[\kappa(\mathcal{R},\mathcal{T}) \,p(r)\Big]=0~~~~\text{or}~~~~\kappa(\mathcal{R},\mathcal{T})\,p(r)=C,
\end{eqnarray}
where $C$ is a constant of integration.

To determine the form of $b(r)$, one needs to insert $df/dr=0$ and $\kappa(\mathcal{R},\mathcal{T})p(r)=C$ into field equations (\ref{array}). We obtain 
\begin{eqnarray}
{b(r) \over r}=-Cr^{2}.
\end{eqnarray}
This solution is not monotonically decreasing, and hence asymptotically non-flat, i.e, it is easy to see that $1-b(r)/r=1+{C}r^{2} \nrightarrow 1$ at $r\to \infty$. The conclusion, contrary to the claim of ref. \cite{Sarkar}, wormhole configurations with constant redshift function are not allowed in $\kappa(\mathcal{R},\mathcal{T})-$ gravity. Our result is in agreement with other modified gravity investigations that also pointed out the non-existence of wormhole solutions for constant redshift function sustained by a non-viscous fluid with isotropic pressure. This in line with the findings of a similar study in $f(\mathcal{R},\mathcal{T})$ gravity theory \cite{Baner}.

\subsection{WH solutions with linear EoS}
The 4 unknowns i.e. $f(r)$, $b(r)$, $p(r)$, $\rho(r)$, and three equations relating them, the system can be fully determined in principle with a linear equation of state $p(r)=\omega \rho(r)$, where $0<\omega<1$. The strategy to solve the equations is the following: from Eq. (\ref{array}), we can express the generic functional $\kappa(\mathcal{R},\mathcal{T})$ in terms of $\rho(r)$ and $b^{\prime}(r)$ as
\begin{eqnarray}
\kappa(\mathcal{R},\mathcal{T})=\frac{b^{\prime}(r)}{r^{2}}\cdot \frac{1}{\rho(r)},~~~~~\forall~~~~\rho(r)\neq 0.
\end{eqnarray}
Hence, inserting this result into Eq. (\ref{nonconservation3}), we obtain the differential equation
\begin{eqnarray}
\omega \rho(r) \displaystyle\frac{d}{dr}\left[\frac{b^{\prime}(r)}{r^{2}}\cdot \frac{1}{\rho(r)}\right]+\frac{b^{\prime}(r)}{r^{2}}\cdot\frac{1}{\rho (r)}\Big[f^{\prime}(r)\rho(r)(1+\omega)\nonumber\\+\omega\rho^{\prime}(r)\Big]=0.~~~~~
\end{eqnarray}
Then, we have at most second order derivatives, namely, higher order derivatives do not appear in our theory, as mentioned before.
After some manipulations this equation leads to
\begin{eqnarray}
f^{\prime}(r)=-\frac{\omega}{\omega+1}\frac{\Big(b^{\prime}(r)/r^{2}\Big)^{\prime}}{b^{\prime}(r)/r^{2}},~~~~~\omega\neq -1,
\end{eqnarray}
which is integrated as
\begin{eqnarray}
f(r)=-\frac{\omega}{\omega+1}\ln \Big(\frac{b^{\prime}(r)}{r^{2}}\Big),~~~~~\omega\neq -1.
\end{eqnarray}
Here, we consider a thin wormhole i.e. $1-\frac{b(r)}{r}<<1$. For such an approximation, we obtain that both field equations can be combined into 

\begin{figure*}[!htb]
\centering
\includegraphics[width=18cm,height=6cm]{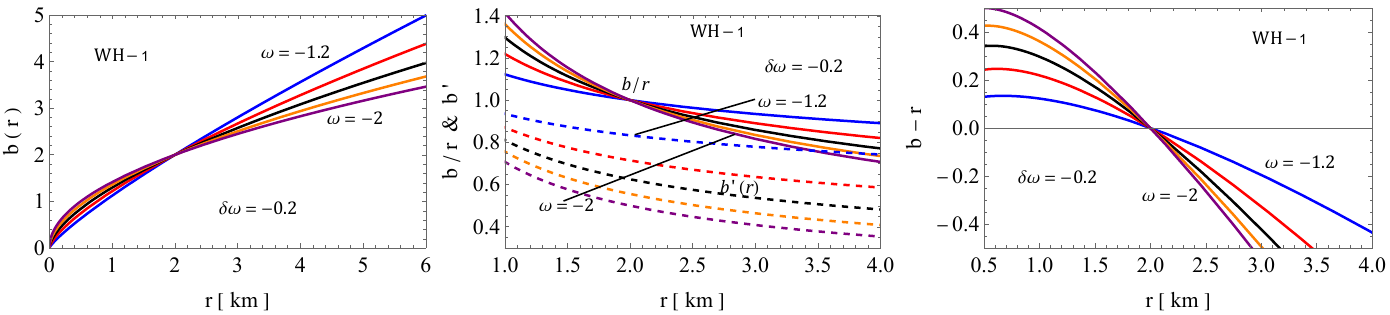}
\caption{Variation of $b(r),~b/r,b-r$ and $b'(r)$ for $r_{_0}=2km$ and $\lambda=0.5$ [WH1].} \label{fig1}
\end{figure*}

\begin{eqnarray}
\omega~ \frac{b^{\prime}(r)}{r^{2}}+\frac{b(r)}{r^{3}}=0,
\end{eqnarray}
from which one obtain $b(r)$ as
\begin{eqnarray}\label{b(r)}
b(r)=r_{_0}\Big(\frac{r_{_0}}{r}\Big)^{1/\omega}~~~~[\text{WH1}].
\end{eqnarray}
the constant of integration was chosen to fulfill the condition $b(r_{_0})=r_{_0}$ at the throat. It is not difficult to show that $b(r)$ satisfies all the stringent constraints to represent a wormhole geometry (see Figs. \ref{fig1}). For instance, when $r=2r_{_0}>r_{_0}$, we have that $b(2r_{_0})/2r_{_0}=(1/2)^{\frac{\omega+1}{\omega}}$, and the condition $b(r)<r$ for $r>r_{_0}$ is therefore satisfied, such as the other requirements. On the other hand, the invariant curvature is given by
\begin{eqnarray}
\mathcal{R}=\frac{2b^{\prime}(r)}{r^{2}}+\frac{d}{dr}\Big(\frac{b(r)}{r}\Big)f^{\prime}(r)=-\frac{3(\omega+1)}{\omega}r_{_0}^{\frac{\omega+1}{\omega}}r^{-\frac{3\omega+1}{\omega}}\nonumber\\\equiv-3(\omega+1)\mu \,r^{-\frac{3\omega+1}{\omega}}.~~~~~
\end{eqnarray}
where it was defined the parameter $\mu=\frac{1}{\omega}r_{_0}^{\frac{\omega+1}{\omega}}$, that depends on $\omega$. Here, this shape-function will only satisfy all the conditions of physical WH iff $\omega<-1$ i.e. supported by phantom energy. Now we proceed to discuss the physical solutions corresponding to different particular choices of the $\kappa(\mathcal{R},\mathcal{T})$ functional.

\begin{figure*}[!htb]
\centering
\includegraphics[width=17cm,height=6cm]{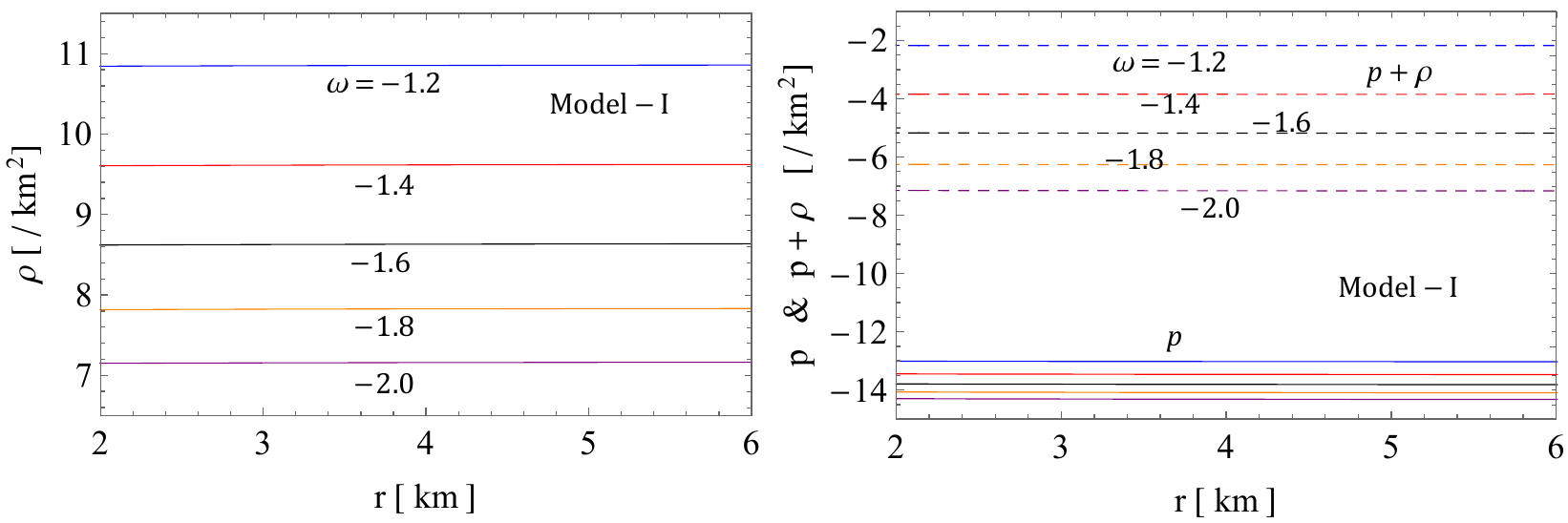}
\caption{Variation of $\rho(r),~p(r)$ and $p+\rho$ for $r_{_0}=2km$ and $\lambda=0.5$ [Model-I].} \label{fig2}
\end{figure*}

\subsubsection{Model I:- $\kappa(\mathcal{T})=8\pi-\lambda \mathcal{T}$ and $b(r)=\Big(\frac{r_{_0}}{r}\Big)^{1/\omega}r_{_0}$}

This particular selection of the $\kappa(\mathcal{R},\mathcal{T})$ running gravitational constant has been subjected to cosmological inspection by several authors. In our case, inserting Eq. (\ref{b(r)}) into the field equations (\ref{array}), we find the following quadratic equation for the density, for $\omega<1/3$ as
\begin{eqnarray}
\rho(r)^{2}-\frac{8\pi}{\lambda(1-3\omega)}\rho(r)-\frac{\mu}{\lambda(1-3\omega)}r^{-\frac{3\omega+1}{\omega}}=0,~~~~~~~\\
\text{or}~~~~\rho(r)=\frac{4\pi}{\lambda(1-3\omega)}\left[1+\sqrt{1+\lambda(1-3\omega)D\, r^{-\frac{3\omega+1}{\omega}}}\right],~~~~~~
\end{eqnarray}
where $D=\mu/16\pi$, and it was selected the positive square root in order to have a consistent physical solution. This density is well behaved for $\omega <1/3$.
\begin{eqnarray}
p(r)=\frac{4\pi \omega}{\lambda(1-3\omega)}\left[1+\sqrt{1+\lambda(1-3\omega)D\, r^{-\frac{3\omega+1}{\omega}}}\right],~~~~~~\\
\rho(r)+p(r)=\frac{4\pi (\omega+1)}{\lambda(1-3\omega)}\left[1+\sqrt{1+\lambda(1-3\omega)D\, r^{-\frac{3\omega+1}{\omega}}}\right].~~~~~~
\end{eqnarray}
The variations of pressure, density and NEC are shown in Fig. \ref{fig2}. Here, the density and pressure are almost constant that also violates the NEC. These results are in agreement with those obtained by Sarkar et al. \cite {Sarkar}, who studied the $\kappa(\mathcal{T})$ case, but under different assumptions. They considered a constant redshift function (the non-conservative equation was not taken into account), while in our case it was considered the thin shell approximation. Remarkably, both cases lead to the same differential equation for the shape function $b(r)$. We proceed to study now the wormhole configurations for other functional forms of Einstein's running gravitational constant.

\begin{figure*}[!htb]
\centering
\includegraphics[width=17cm,height=6cm]{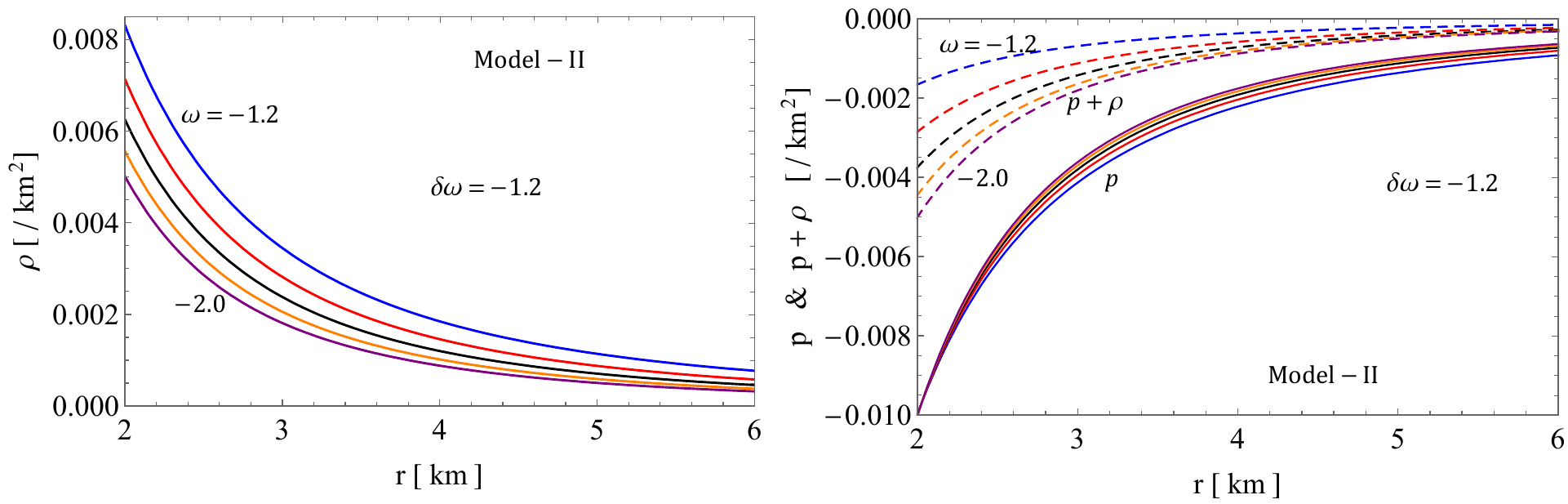}
\caption{Variation of $\rho(r),~p(r)$ and $p+\rho$ for $r_{_0}=2km$ and $\beta=0.5$ [Model-II].} \label{fig3}
\end{figure*}

\subsubsection{Model II:- $\kappa(\mathcal{R})=\beta \mathcal{R}+8\pi$ and $b(r)=\Big(\frac{r_{_0}}{r}\Big)^{1/\omega}r_{_0}$}
The density and isotropic pressure for this particular choice of the $\kappa(\mathcal{R},\mathcal{T})=8\pi+\beta \mathcal{R}$ functional can be also obtained by analytic means. This time the algebraic equation for the density is not quadratic. Indeed, the substitution of $\kappa(\mathcal{R},\mathcal{T})=8\pi+\beta \mathcal{R}$ into Eq. (\ref{array}) leads to
\begin{eqnarray}
\left[8\pi-C\mu \,r^{-\frac{3\omega+1}{\omega}}\right]\rho(r)=-\mu \,r^{-\frac{3\omega+1}{\omega}},
\end{eqnarray}
with $C=3\beta(1+\omega)$.
\begin{eqnarray}
\rho(r) &=& \frac{1}{C-E\,r^{\frac{3\omega+1}{\omega}}},\\
p(r)&=&\frac{\omega}{C-E\,r^{\frac{3\omega+1}{\omega}}},\\
p(r)+\rho(r)&=&\frac{\omega+1}{C-E\,r^{\frac{3\omega+1}{\omega}}},
\end{eqnarray}
where $E=8\pi/\mu$.
The variations of pressure, density and NEC are shown in Fig. \ref{fig3}. Here, the density and pressure decreases with radial distance away from the throat along with the violation of the NEC.

\begin{figure*}[!htb]
\centering
\includegraphics[width=17cm,height=6cm]{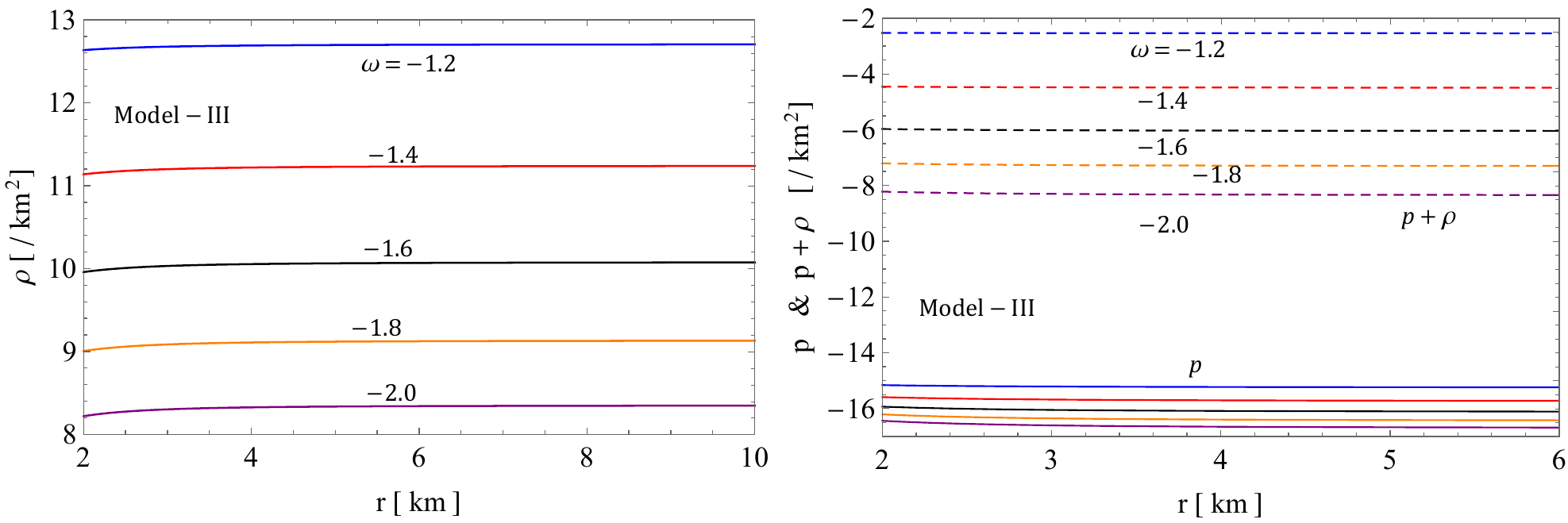}
\caption{Variation of $\rho(r),~p(r)$ and $p+\rho$ for $r_{_0}=2km,~\beta=0.5$ and $\lambda=0.43$ [Model-III].} \label{fig4}
\end{figure*}

\subsubsection{Model III:- $\kappa(\mathcal{R},\mathcal{T})=\beta\mathcal{R}-\lambda \mathcal{T}+8\pi$ and $b(r)=\Big(\frac{r_{_0}}{r}\Big)^{1/\omega}r_{_0}$}

For this specific case, the density and pressure are given by the following expressions
\begin{eqnarray}
\rho(r)&=&\frac{4\pi-C\mu\, r^{-\frac{3\omega+1}{\omega}}}{\lambda(1-3\omega)}\nonumber\\
&& \times\left[1+\sqrt{1+4\mu(1-3\omega)\lambda
\psi(r)}\right],~~~~~~~
\end{eqnarray}
where, like in the above equations, $C=3\beta(1+\omega)$, $\mu=\frac{1}{\omega}r_{_0}^{\frac{\omega+1}{\omega}}$, and we have defined certain function $\psi(r)$ as
\begin{eqnarray}
\psi(r)&=&\frac{r^{-\frac{3\omega+1}{\omega}}}{\left[8\pi-C\mu \,r^{-\frac{3\omega+1}{\omega}}\right]^{2}},
\end{eqnarray}
\begin{eqnarray}
p(r)&=&\frac{\omega(4\pi-C\mu \,r^{-\frac{3\omega+1}{\omega}})}{\lambda(1-3\omega)}\nonumber\\
&& \times\left[1+\sqrt{1+4\mu(1-3\omega)\lambda\psi(r)}\right],\\
p(r)+\rho(r)&=&\frac{(\omega+1)\big(4\pi-C\mu r^{-\frac{3\omega+1}{\omega}}\big)}{\lambda(1-3\omega)}\nonumber\\
&& \times \left[1+\sqrt{1+4\mu(1-3\omega)\lambda
\psi(r)}\right].
\end{eqnarray}
The variations of pressure, density and NEC are shown in Fig. \ref{fig4}.Again, here the pressure and density are almost constant and also $\rho+p<0$ implying the violation of NEC.

\begin{figure*}[!htb]
\centering
\includegraphics[width=17cm,height=6cm]{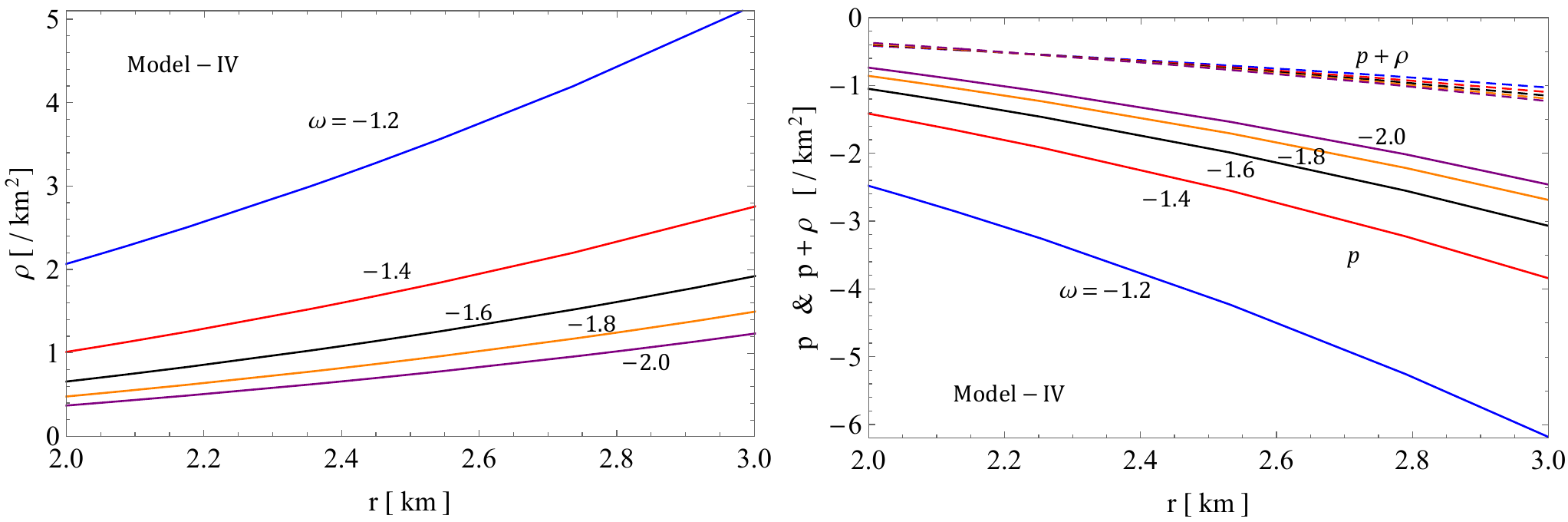}
\caption{Variation of $\rho(r),~p(r)$ and $p+\rho$ for $r_{_0}=2km,$ and $\gamma=-20$ [Model-IV].} \label{fig5}
\end{figure*}
\subsubsection{Model IV:- $\kappa(\mathcal{R},\mathcal{T})=8\pi-\gamma\mathcal{R}\mathcal{T}$ and $b(r)=\Big(\frac{r_{_0}}{r}\Big)^{1/\omega}r_{_0}$}
This choice of the $\kappa(\mathcal{R},\mathcal{T})=8\pi-\gamma \mathcal{R}\mathcal{T}$ was studied in ref.\cite{Ter2}, for gravastar configurations. After a bit of algebra, we obtain:
\begin{eqnarray}
\rho(r)&=&\frac{-4\pi r^{\frac{3\omega+1}{\omega}}}{3\mu\gamma(1-3\omega)(\omega+1)}\nonumber\\&& \times \left[1+\sqrt{1-a(1-3\omega)(\omega+1)r^{-\frac{3\omega+1}{\omega}}}\right].
\end{eqnarray}
where we have defined the constant $a=\frac{9\gamma\mu^{2}}{16\pi^{2}}$, and the parameter $\mu$ that depends on $\omega$, is the same that was defined above. The density is positive for $\omega>1/3$ and $\omega<-1$. However, the first choice $\omega>1/3$ leads to a non-asymptotically flat spacetime along with the violation of $b'(r_0)>0$, hence the only possible choice is the second one i.e. $\omega<-1$ for which the pressure is negative while the density is positive and increasing function of $r$. On the other hand, the pressure and the sum of both quantities, density and pressure, are given by
\begin{eqnarray}
p(r)&=&\frac{-4\pi \omega r^{\frac{3\omega+1}{\omega}}}{3\mu\gamma(1-3\omega)(\omega+1)}\nonumber\\&&\hspace{-0.3cm} \times\left[1+\sqrt{1-a(1-3\omega)(\omega+1)r^{-\frac{3\omega+1}{\omega}}}\right],\\
\rho(r)+p(r)&=&\frac{-4\pi (\omega+1) r^{\frac{3\omega+1}{\omega}}}{3\mu\gamma(1-3\omega)(\omega+1)}\nonumber\\&&\hspace{-0.3cm} \times\left[1+\sqrt{1-a(1-3\omega)(\omega+1)r^{-\frac{3\omega+1}{\omega}}}\right].~~~~
\end{eqnarray}
The trend of pressure, density and NEC are shown in Fig. \ref{fig5}. Here the pressure and density are non-decreasing when move away from the throat and $\rho+p<0$, so again violates the NEC.

\section{Wormhole Configurations for Specific Choices of the Shape Function}
We have four unknown functions $f(r)$, $b(r)$, $\rho(r)$, $p(r)$, and three equations linking them, two field equations and the non-conservative equation. Therefore, we have one degree of freedom. In the last subsection, we completely solved the system by assuming a linear EoS, $p(r)=\omega \rho(r)$, which is enough to solve the equations. Here, we consider that the relation between $p(r)$ and $\rho(r)$ is not linear. Therefore, our aim is to obtain a differential equation between $b(r)$ and $f(r)$. Then, assume one form for the shape function satisfying the wormhole conditions, and after computing $f(r)$, finally the density and pressure can be computed. Recall that the curvature scalar depends also on $b(r)$ and $f^{\prime}(r)$. From equations (\ref{array}), we obtain
\begin{eqnarray}
\frac{\rho(r)}{p(r)}=\frac{\frac{b^{\prime}(r)}{r^{2}}}{2\frac{f^{\prime}(r)}{r}\left(1-\frac{b(r)}{r}\right)-\frac{b(r)}{r^{3}}}~~ \cdot
\end{eqnarray}
Like in the section before, we adopt the approximation $1-\frac{b(r)}{r}<<1$, which means that we consider a thin wormhole, so that the dimensions of these structures are not arbitrarily large. Then
\begin{eqnarray}
\frac{\rho(r)}{p(r)}=-\frac{r\,b^{\prime}(r)}{b(r)}.
\end{eqnarray}

\begin{figure*}[!htb]
\centering
\includegraphics[width=18cm,height=6cm]{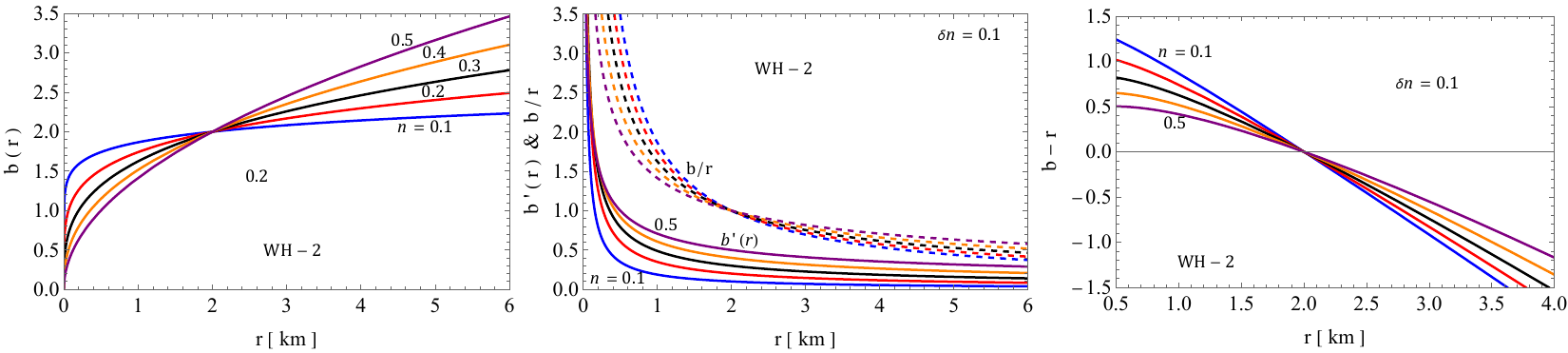}
\caption{Variation of $b(r),~b/r,b-r$ and $b'(r)$ for $r_{_0}=2km$ and $\lambda=0.5$ [WH2].} \label{fig6}
\end{figure*}

Inserting this result together with $\kappa(\mathcal{R},\mathcal{T})=\frac{b^{\prime}(r)}{r^{2}}\cdot \frac{1}{\rho(r)}$ into the non-conservative equation, we obtain the differential equation
\begin{eqnarray}
\frac{d}{dr}\left(\frac{b(r)}{r^{3}}\right)+\frac{b(r)}{r^{3}}
\left[f^{\prime}(r)\Big(1-\frac{b^{\prime}(r)}{b(r)}\cdot r\Big)\right]=0 .\label{e38}
\end{eqnarray}
By choosing specific form of the shape or redshift functions, one can solve Eq. \eqref{e38} and test their physical acceptability conditions. We have chosen few of the well-known shape functions to check existence of any physically inspired WH solutions in different forms of $\kappa(\mathcal{R},\mathcal{T})-$gravity.

\subsection{WH with $b(r)=\left(r/\beta \right) ^{n} \beta$ ~~[\text{WH2}]}
Using this shape function in \eqref{e38}, we get the redshift function as
\begin{eqnarray}
f \left( r \right) =\left(\frac{n-3 }{n-1}\right) \ln r+c_1,
\end{eqnarray}
with $c_1$ is the constant of integration. The variation of the shaped function, $b'(r)$, $b/r$ and $b-r$ are shown in Fig. \ref{fig6}. This form of shaped-function can hold all the properties of a WH when $n$ is chosen between the interval $]0,1[$.

\subsubsection{Model V:- $\kappa(\mathcal{T})=8\pi-\lambda \mathcal{T}$ and $b(r)=\left( {\frac {r}{\beta}} \right)^{n}\beta$}
Now, the density and pressure takes the form,
\begin{widetext}
\begin{eqnarray}
\rho &=& \frac{\beta \lambda^{1/2} (n-1) n \left(\frac{r}{\beta }\right)^n \left[\beta  \left(n^2+5 n-18\right) \left(\frac{r}{\beta }\right)^n+3 \beta  (n-1) r^2 \left(\frac{r}{\beta }\right)^n-6 (n-3) r\right]^{-1} }{\beta  (n-1) r^2 \left(\frac{r}{\beta }\right)^n+2 \beta  (n-3) \left(\frac{r}{\beta }\right)^n-2 (n-3) r} \nonumber \\
&& \bigg[4 \pi  \beta  n r^2 \left(\frac{r}{\beta }\right)^n-4 \pi  \beta  r^2 \left(\frac{r}{\beta }\right)^n+8 \pi  \beta  n \left(\frac{r}{\beta }\right)^n-24 \pi  \beta  \left(\frac{r}{\beta }\right)^n-8 \pi  n r+24 \pi  r \bigg] \nonumber \\
&& +\left[\beta  (n-1) r^2 \left(\frac{r}{\beta }\right)^n+2 \beta  (n-3) \left(\frac{r}{\beta }\right)^n-2 (n-3) r\right] \times \nonumber \\
&& \hspace{-0.5cm} \sqrt{\frac{16 \pi ^2 (n-1) r^3-\lambda  \left[\beta  \left(n^2+5 n-18\right) \left(\frac{r}{\beta }\right)^n+3 \beta  (n-1) r^2 \left(\frac{r}{\beta }\right)^n-6 (n-3) r\right]}{(n-1) r^3}},\\
p &=& -\bigg[4 \pi  \beta  n r^2 \left(\frac{r}{\beta }\right)^n-4 \pi  \beta  r^2 \left(\frac{r}{\beta }\right)^n+8 \pi  \beta  n \left(\frac{r}{\beta }\right)^n-24 \pi  \beta  \left(\frac{r}{\beta }\right)^n-8 \pi  n r+24 \pi  r \nonumber \\
&& +\lambda^{1/2}\left(\beta  (n-1) r^2 \left(\frac{r}{\beta }\right)^n+2 \beta  (n-3) \left(\frac{r}{\beta }\right)^n-2 (n-3) r\right) \times \nonumber \\
&& \hspace{-0.5cm} \sqrt{\frac{16 \pi ^2 (n-1) r^3-\left(\beta  \left(n^2+5 n-18\right) \left(\frac{r}{\beta }\right)^n+3 \beta  (n-1) r^2 \left(\frac{r}{\beta }\right)^n-6 (n-3) r\right)}{(n-1) r^3}}\bigg]\nonumber \\
&& \times ~{\lambda^{-1} \over \left(\beta  \left(n^2+5 n-18\right) \left(\frac{r}{\beta }\right)^n+3 \beta  (n-1) r^2 \left(\frac{r}{\beta }\right)^n-6 (n-3) r\right)}.
\end{eqnarray}
\end{widetext}

\begin{figure*}[!htb]
\centering
\includegraphics[width=17cm,height=6cm]{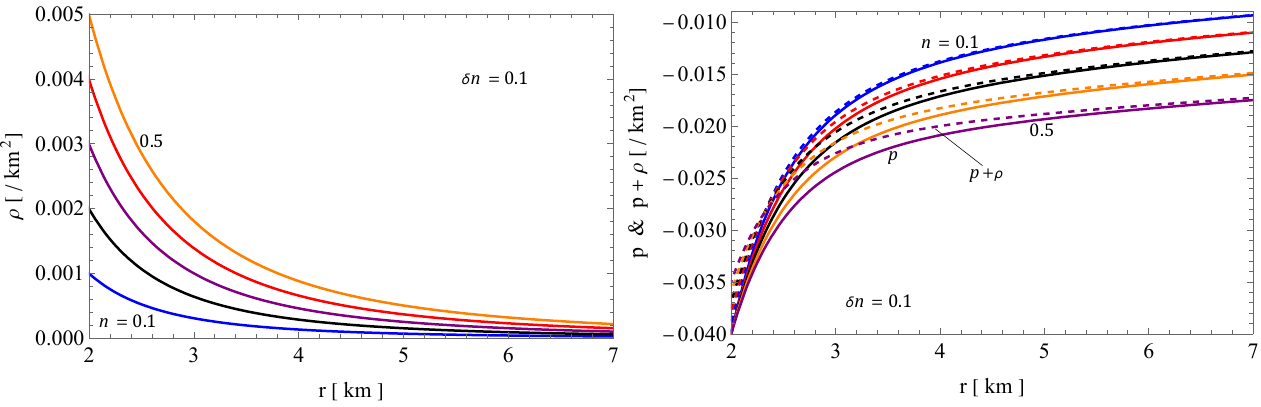}
\caption{Variation of $\rho(r),~p(r)$ and $p+\rho$ for $r_{_0}=2km$ and $\lambda=0.5$ [Model-V].} \label{fig7}
\end{figure*}

The variations of pressure, density and NEC are shown in Fig. \ref{fig7}. This WH has decreasing trends of density and pressure with the NEC still violates.

\subsubsection{Model VI:- $\kappa(\mathcal{R},\mathcal{T})=\xi \mathcal{R}-\lambda \mathcal{T}+8\pi$ and $b(r)=\left( {\frac {r}{\beta}} \right)^{n}\beta$}
For this form of $\kappa(\mathcal{R},\mathcal{T})$ function, the density and pressure takes the form
\begin{widetext}
\begin{eqnarray}
\rho &=& \frac{\beta n \sqrt{n-1} \left(\frac{r}{\beta }\right)^n}{2 \lambda  r^3 \left(\beta  \left(n^2+5 n-18\right) \left(\frac{r}{\beta }\right)^n+3 \beta  (n-1) r^2 \left(\frac{r}{\beta }\right)^n-6 (n-3) r\right)} \bigg[8 \pi  \sqrt{n-1} ~r^3- \nonumber \\
&& \Big[-4 \beta  \lambda  \left(n^2+5 n-18\right) r^3 \left(\frac{r}{\beta }\right)^n+64 \pi ^2 (n-1) r^6-12 \beta  \lambda  (n-1) r^5 \left(\frac{r}{\beta }\right)^n+ \nonumber \\
&& 24 \lambda  (n-3) r^4+48 \pi  \beta  (n-1)^2 \xi  r^3 \left(\frac{r}{\beta }\right)^n+9 \beta ^2 (n-1)^3 \xi ^2 \left(\frac{r}{\beta }\right)^{2 n}\Big]^{1/2}- \nonumber \\
&& 3 \beta \xi \sqrt{n-1}  \left(\frac{r}{\beta }\right)^n+3 \beta  n \xi  \sqrt{n-1} \left(\frac{r}{\beta }\right)^n \bigg],\\
p &=& -\frac{\beta  (n-1) r^2 \left(\frac{r}{\beta }\right)^n+2 \beta  (n-3) \left(\frac{r}{\beta }\right)^n+2 (3-n) r}{2 \lambda  \sqrt{n-1} ~r^3 \left(\beta  \left(n^2+5 n-18\right) \left(\frac{r}{\beta }\right)^n+3 \beta  (n-1) r^2 \left(\frac{r}{\beta }\right)^n-6 (n-3) r\right)} \nonumber \\
&& \bigg[8 \pi  \sqrt{n-1}~ r^3-3 \beta  \sqrt{n-1}~ \xi  \left(\frac{r}{\beta }\right)^n+3 \beta  n \xi  \sqrt{n-1} \left(\frac{r}{\beta }\right)^n -\Big\{-4 \beta  \lambda  \left(n^2+5 n-18\right) r^3 \nonumber \\
&& \left(\frac{r}{\beta }\right)^n+64 \pi ^2 (n-1) r^6-12 \beta  \lambda  (n-1) r^5 \left(\frac{r}{\beta }\right)^n+24 \lambda  (n-3) r^4+48 \pi  \beta  (n-1)^2 \xi  r^3 \left(\frac{r}{\beta }\right)^n \nonumber \\
&& +9 \beta ^2 (n-1)^3 \xi ^2 \left(\frac{r}{\beta }\right)^{2 n} \Big\}^{1/2}\bigg].
\end{eqnarray}
\end{widetext}
The variations of pressure, density and NEC are exactly same as in Model V as well as Model VI (Fig. \ref{fig7}).
\begin{figure*}[!htb]
\centering
\includegraphics[width=18cm,height=6cm]{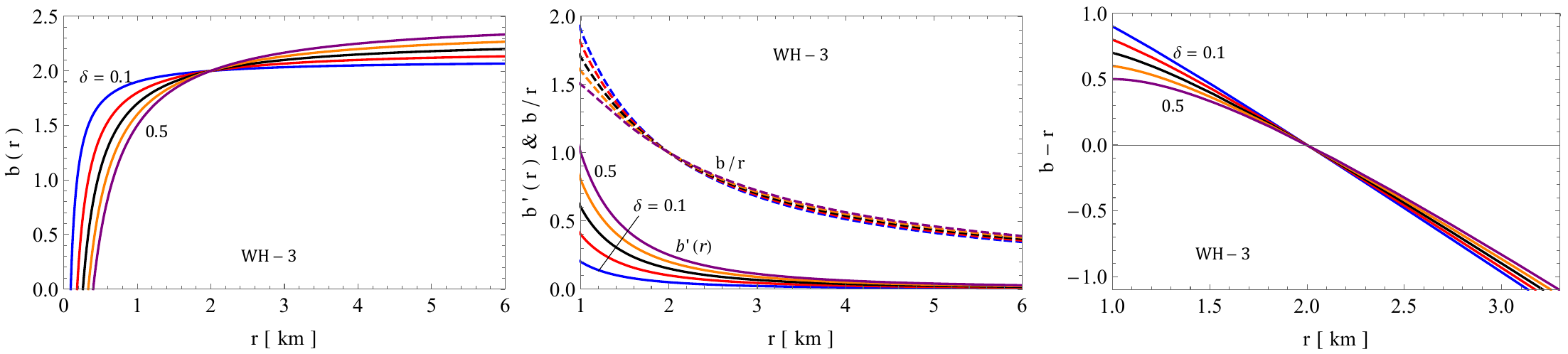}
\caption{Variation of $b(r),~b/r,b-r$ and $b'(r)$ for $a=2km,~\xi =0.3$ and $\lambda=2$ [WH3].} \label{fig8}
\end{figure*}

\subsubsection{Model VII:- $\kappa(\mathcal{R},\mathcal{T})=8\pi-\gamma \mathcal{R} \mathcal{T}$ and $b(r)=\left( {\frac {r}{\beta}} \right)^{n}\beta$}
Here, the $\kappa(\mathcal{R},\mathcal{T})$ running gravitational constant represents a direct coupling between matter and geometry, creating a nonlinear functional form with density and pressure as follows:
\begin{widetext}
\begin{small}
\begin{eqnarray}
\rho &=& \frac{n \left(\beta  \left(n^2+5 n-18\right) \left(\frac{r}{\beta }\right)^n+3 \beta  (n-1) r^2 \left(\frac{r}{\beta }\right)^n-6 (n-3) r\right)^{-1}}{3 \gamma  \left(\beta  (n-1) r^2 \left(\frac{r}{\beta }\right)^n+2 \beta  (n-3) \left(\frac{r}{\beta }\right)^n-2 (n-3) r\right)} \bigg[- n \nonumber\\
&& \Bigg\{\frac{(n-1)^{-2}\left(\beta  (n-1) r^2 \left(\frac{r}{\beta }\right)^n+2 \beta  (n-3) \left(\frac{r}{\beta }\right)^n-2 (n-3) r\right)^2}{\left(16 \pi ^2 r^6-3 \beta  \gamma  \left(\frac{r}{\beta }\right)^n \left(\beta  \left(n^2+5 n-18\right) \left(\frac{r}{\beta }\right)^n+3 \beta  (n-1) r^2 \left(\frac{r}{\beta }\right)^n-6 (n-3) r\right)\right)^{-1}} \Bigg\}^{1/2} \nonumber \\
&& + \Bigg\{\frac{(n-1)^{-2}\left(\beta  (n-1) r^2 \left(\frac{r}{\beta }\right)^n+2 \beta  (n-3) \left(\frac{r}{\beta }\right)^n-2 (n-3) r\right)^2 }{ \left(16 \pi ^2 r^6-3 \beta  \gamma  \left(\frac{r}{\beta }\right)^n \left(\beta  \left(n^2+5 n-18\right) \left(\frac{r}{\beta }\right)^n+3 \beta  (n-1) r^2 \left(\frac{r}{\beta }\right)^n-6 (n-3) r\right)\right)^{-1}} \Bigg\}^{1/2} \nonumber \\
&& +4 \pi  \beta  n r^5 \left(\frac{r}{\beta }\right)^n-4 \pi  \beta  r^5 \left(\frac{r}{\beta }\right)^n-8 \pi  n r^4+8 \pi  \beta  n r^3 \left(\frac{r}{\beta }\right)^n-24 \pi  \beta  r^3 \left(\frac{r}{\beta }\right)^n+24 \pi  r^4 \bigg]\\
p &=& \frac{\left(r/\beta\right)^{-n}}{3 \beta  \gamma  (n-1) \left(\beta  \left(n^2+5 n-18\right) \left(\frac{r}{\beta }\right)^n+3 \beta  (n-1) r^2 \left(\frac{r}{\beta }\right)^n-6 (n-3) r\right)} \nonumber \\
&& \Bigg[4 \pi  \beta  r^5 \left(\frac{r}{\beta }\right)^n-4 \pi  \beta  n r^5 \left(\frac{r}{\beta }\right)^n+8 \pi  n r^4-8 \pi  \beta  n r^3 \left(\frac{r}{\beta }\right)^n+24 \pi  \beta  r^3 \left(\frac{r}{\beta }\right)^n-24 \pi  r^4 \nonumber \\
&& -\sqrt{\frac{\left(16 \pi ^2 r^6-3 \beta  \gamma  \left(\frac{r}{\beta }\right)^n \left(\beta  \left(n^2+5 n-18\right) \left(\frac{r}{\beta }\right)^n+3 \beta  (n-1) r^2 \left(\frac{r}{\beta }\right)^n-6 (n-3) r\right)\right)}{(n-1)^2 \left(\beta  (n-1) r^2 \left(\frac{r}{\beta }\right)^n+2 \beta  (n-3) \left(\frac{r}{\beta }\right)^n-2 (n-3) r\right)^{-2}}} \nonumber \\
&& \hspace{-0.5cm} +n \sqrt{\frac{ \left(16 \pi ^2 r^6-3 \beta  \gamma  \left(\frac{r}{\beta }\right)^n \left(\beta  \left(n^2+5 n-18\right) \left(\frac{r}{\beta }\right)^n+3 \beta  (n-1) r^2 \left(\frac{r}{\beta }\right)^n-6 (n-3) r\right)\right)}{(n-1)^2 \left(\beta  (n-1) r^2 \left(\frac{r}{\beta }\right)^n+2 \beta  (n-3) \left(\frac{r}{\beta }\right)^n-2 (n-3) r\right)^{-2}}}\Bigg].
\end{eqnarray}
\end{small}
\end{widetext}
The variations of pressure, density and NEC are exactly same as in Model V Fig. \ref{fig7}. This model also have the exact same density and pressure profile for the same set of parameter with Model-V and VI.

\subsection{WH with $b(r)=a +\delta  \left(1-\frac{a}{r}\right)$ [WH3]}
This choice shape functions leads to the redshift function as,
\begin{eqnarray}
f(r)=c_1+\ln (a r+\delta  r -2 a \delta )+2 \ln r~.
\end{eqnarray}
Now for specific form of $\kappa(\mathcal{R},\mathcal{T})$ we can determine the density and pressure. The variation of the shaped function, $b'(r)$, $b/r$ and $b-r$ are shown in Fig. \ref{fig8}.

\subsubsection{Model VIII:- $\kappa(\mathcal{R},\mathcal{T})=\xi \mathcal{R}-\lambda \mathcal{T}+8\pi$ and $b(r)=a +\delta  \left(1-\frac{a}{r}\right)$}
The density and pressure for this model is given by 

\begin{figure*}
\centering
\includegraphics[width=16cm,height=6.5cm]{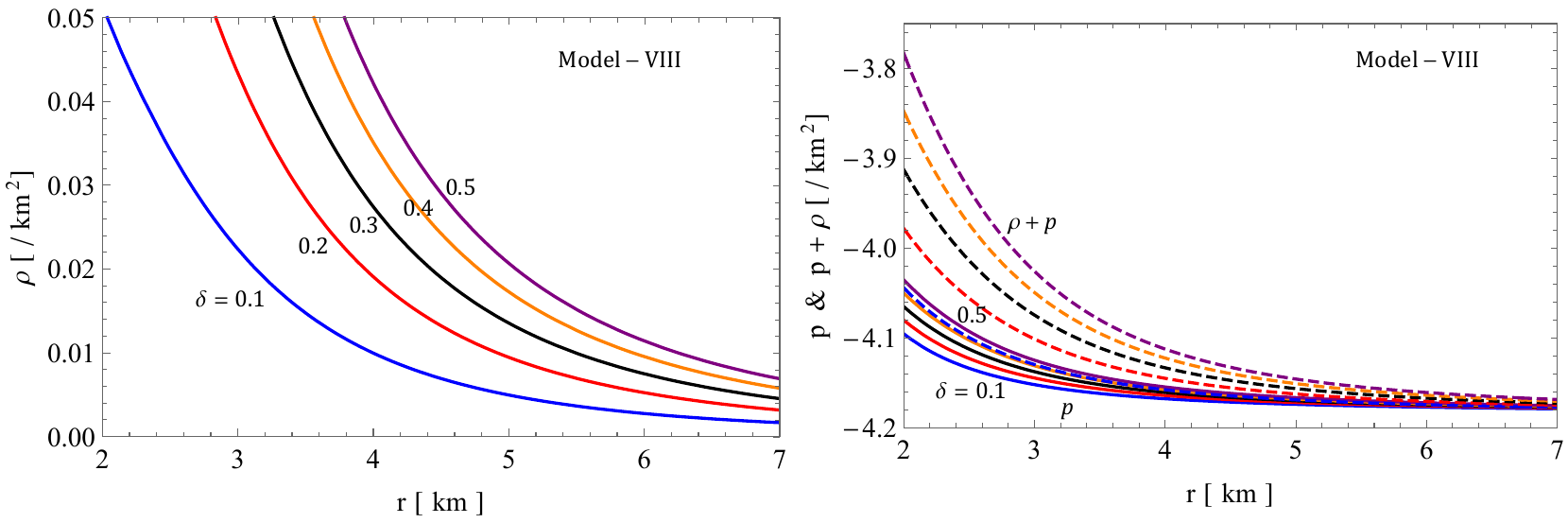}
\caption{Variation of $\rho(r),~p(r)$ and $p+\rho$ for $a=2km,~\xi =0.3$ and $\lambda=2$ [Model-VIII].} \label{fig9}
\end{figure*}

\begin{widetext}
\begin{small}
\begin{eqnarray}
\rho &=& \frac{a \delta }{2 \lambda  r^4} \Big[\Big\{a^2 (r-\delta ) \left(r \left(r^2+6\right)-2 \delta  \left(r^2+4\right)\right)+a r \left(-14 \delta ^2+2 \delta  r^3-3 \left(\delta ^2+2\right) r^2+20 \delta  r\right)+\delta  r^2 \nonumber\\ 
&& \hspace{-0.7cm} \left(\delta  \left(r^2+6\right)-6 r\right)\Big\} \Big\{a^2 \left(22 \delta ^2+3 r^4-9 \delta  r^3+6 \left(\delta ^2+3\right) r^2-41 \delta  r\right)+a r (-41 \delta ^2+6 \delta  r^3-9  \nonumber\\
&&  \hspace{-0.7cm} \left(\delta ^2+2\right) r^2+60 \delta  r)+3 \delta  r^2 \left(\delta  \left(r^2+6\right)-6 r\right)\Big\} \Big]^{-1} \Big[\Big\{(a (r-2 \delta )+\delta  r) (a^2 (r-\delta ) (r \left(r^2+6\right)-2 \delta  \nonumber\\
&& \hspace{-0.7cm} \left(r^2+4\right))+a r \left(-14 \delta ^2+2 \delta  r^3-3 \left(\delta ^2+2\right) r^2+20 \delta  r\right)+\delta  r^2 \left(\delta  \left(r^2+6\right)-6 r\right))^2 (-4 \lambda  r^4 (a^2 \nonumber\\
&& \hspace{-0.7cm} \left(22 \delta ^2+3 r^4-9 \delta  r^3+6 \left(\delta ^2+3\right) r^2-41 \delta  r\right)+a r \left(-41 \delta ^2+6 \delta  r^3-9 \left(\delta ^2+2\right) r^2+60 \delta  r\right)+3 \delta  r^2 \nonumber\\
&& \hspace{-0.7cm} \left(\delta  \left(r^2+6\right)-6 r\right))+64 \pi ^2 r^8 (a (r-2 \delta )+\delta  r)-48 \pi  \xi  r^4 (a (r-2 \delta )+\delta  r)^2+9 \xi ^2 (a (r-2 \delta )+\delta  r)^3)\Big\}^{1/2} \nonumber\\
&& \hspace{-0.7cm} -(a (r-2 \delta )+\delta  r) (a^2 (r-\delta ) \left(r \left(r^2+6\right)-2 \delta  \left(r^2+4\right)\right)+a r \left(-14 \delta ^2+2 \delta  r^3-3 \left(\delta ^2+2\right) r^2+20 \delta  r\right) \nonumber\\
&& \hspace{-0.7cm} +\delta  r^2 \left(\delta  \left(r^2+6\right)-6 r\right)) \left(3 \xi  (a (r-2 \delta )+\delta  r)-8 \pi  r^4\right) \Big], \\
p &=& \Big[(a (r-2 \delta )+\delta  r) (a^2 (r-\delta ) \left(r \left(r^2+6\right)-2 \delta  \left(r^2+4\right)\right)+a r \left(-14 \delta ^2+2 \delta  r^3-3 \left(\delta ^2+2\right) r^2+20 \delta  r\right) \nonumber\\
&& \hspace{-0.7cm} +\delta  r^2 \left(\delta  \left(r^2+6\right)-6 r\right)) \left(3 \xi(a (r-2 \delta )+\delta  r)-8 \pi  r^4\right)-\Big\{(a (r-2 \delta )+\delta  r) (a^2 (r-\delta ) (r \left(r^2+6\right)-2 \delta  \nonumber\\
&& \hspace{-0.7cm} \left(r^2+4\right))+a r \left(-14 \delta ^2+2 \delta  r^3-3 \left(\delta ^2+2\right) r^2+20 \delta  r\right)+\delta  r^2 \left(\delta  \left(r^2+6\right)-6 r\right))^2 (-4 \lambda  r^4 (a^2(22 \delta ^2 \nonumber\\
&& \hspace{-0.7cm} +3 r^4-9 \delta  r^3+6 \left(\delta ^2+3\right) r^2-41 \delta  r)+a r \left(-41 \delta ^2+6 \delta  r^3-9 \left(\delta ^2+2\right) r^2+60 \delta  r\right)+3 \delta  r^2 \nonumber\\
&& \hspace{-0.7cm} \left(\delta  \left(r^2+6\right)-6 r\right))+64 \pi ^2 r^8 (a (r-2 \delta )+\delta  r)-48 \pi  \xi  r^4 (a (r-2 \delta )+\delta  r)^2+9 \xi ^2 (a (r-2 \delta )+\delta  r)^3)\Big\}^{1/2} \Big] \nonumber \\
&& \hspace{-0.7cm} \Big[2 \lambda  r^4 (a (r-2 \delta )+\delta  r) (a^2 \left(22 \delta ^2+3 r^4-9 \delta  r^3+6 \left(\delta ^2+3\right) r^2-41 \delta  r\right)+a r (-41 \delta ^2+6 \delta  r^3- \nonumber \\
&& \hspace{-0.7cm} 9 \left(\delta ^2+2\right) r^2+60 \delta  r)+3 \delta  r^2 \left(\delta  \left(r^2+6\right)-6 r\right)) \Big]^{-1}
\end{eqnarray}
\end{small}
\end{widetext}
Variations of the pressure and density are shown in Fig. \ref{fig9}.

\subsection{WH with $b(r)=a +a^3 \delta  \ln \left(\gamma /r\right)$ [WH4]}
\begin{figure*}
\centering
\includegraphics[width=17cm,height=6.5cm]{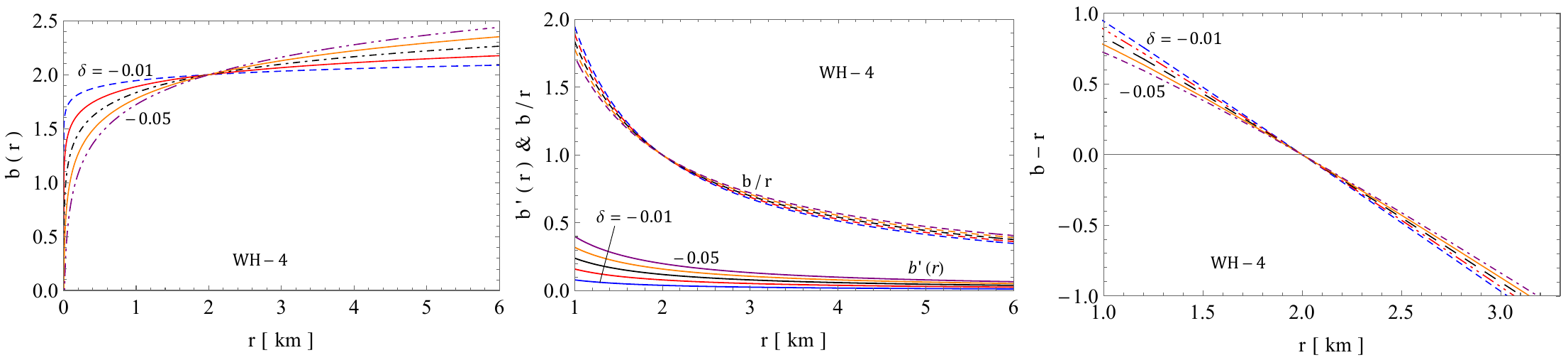}
\caption{Variation of $b(r),~b/r,b-r$ and $b'(r)$ for $a=2km,~\xi =0.3$ and $\lambda=2$ [WH4].} \label{fig10}
\end{figure*}

This choice leads to the redshift function as, 
\begin{eqnarray}
f(r)=2 \log \left[a^2 \delta +a^2 \delta  \ln \left(\frac{a}{r}\right)+1\right]-\frac{3}{a^2 \delta }\nonumber\\-3 \ln \left(\frac{a}{r}\right)+c_1-3.~~~~~
\end{eqnarray}
The variation of the shaped function, $b'(r)$, $b/r$ and $b-r$ are shown in Fig. \ref{fig10}.

\subsubsection{Model IX:- $\kappa(\mathcal{R})=\xi \mathcal{R}+8\pi$ and $b(r)=a +a^3 \delta  \ln \left(\gamma/r\right)$ }
\begin{figure*}
\centering
\includegraphics[width=17cm,height=7cm]{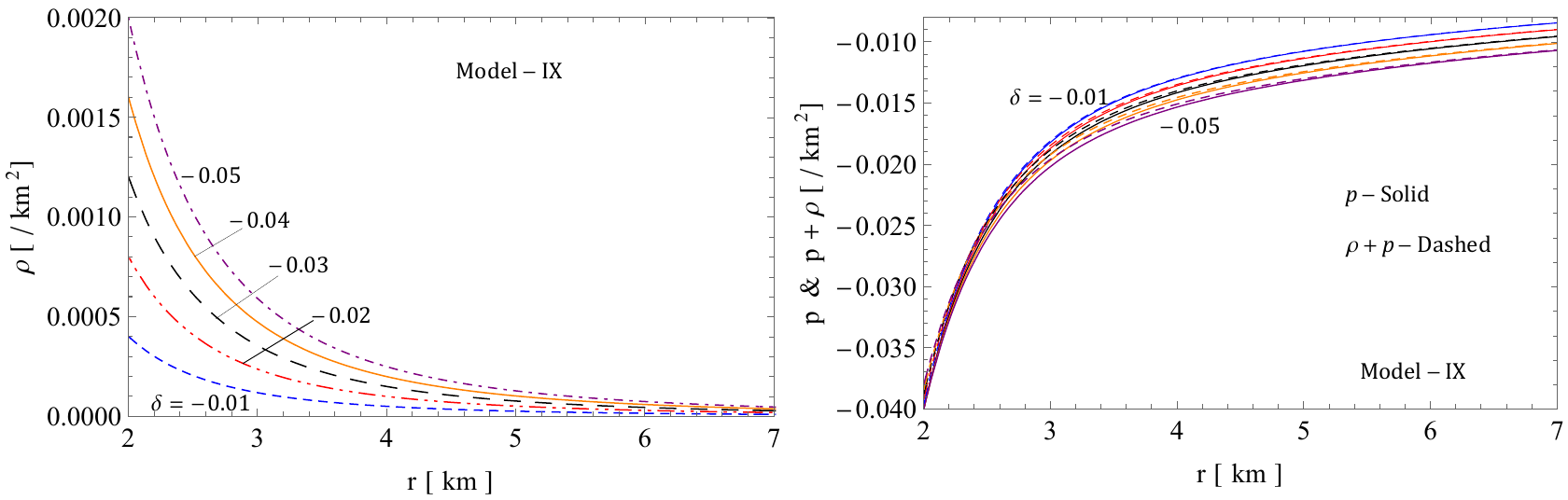}
\caption{Variation of $\rho(r),~p(r)$ and $p+\rho$ for $a=2km,~\xi =0.3$ and $\lambda=2$ [Model-IX].} \label{fig11}
\end{figure*}

This model leads to 
\begin{eqnarray}
\rho &=& \frac{a^3 \delta }{3 a^3 \delta  \xi  \ln \left(\frac{a}{r}\right)+3 a \xi  \left(a^2 \delta +1\right)-8 \pi  r^3}\\
p &=& \Big[a^2 \delta  \ln \left(\frac{a}{r}\right) \Big\{a^3 \delta  \left(r^2+2\right)+a^3 \delta  \left(r^2+6\right) \ln \left(\frac{a}{r}\right)\nonumber \\
&&+2 a \left(r^2+6\right)-6 r\Big\}+a  (a \delta  \left(a \left(r^2+2\right)-2 r\right)+r^2\nonumber \\
&&+6)-6 r \Big] \left(a^2 \delta +a^2 \delta  \log \left(\frac{a}{r}\right)+1\right)^{-1}  \bigg(3 a^3 \delta  \xi \nonumber \\
&& \times  \log \left(\frac{a}{r}\right)+3 a \xi  \left(a^2 \delta +1\right)-8 \pi  r^3\bigg)^{-1}. 
\end{eqnarray}
Variations of the pressure and density are shown in Fig. \ref{fig11}.

\section{Calculation of embedding surface and the proper radial distance}

\begin{figure*}
\centering
\includegraphics[width=7cm,height=7cm]{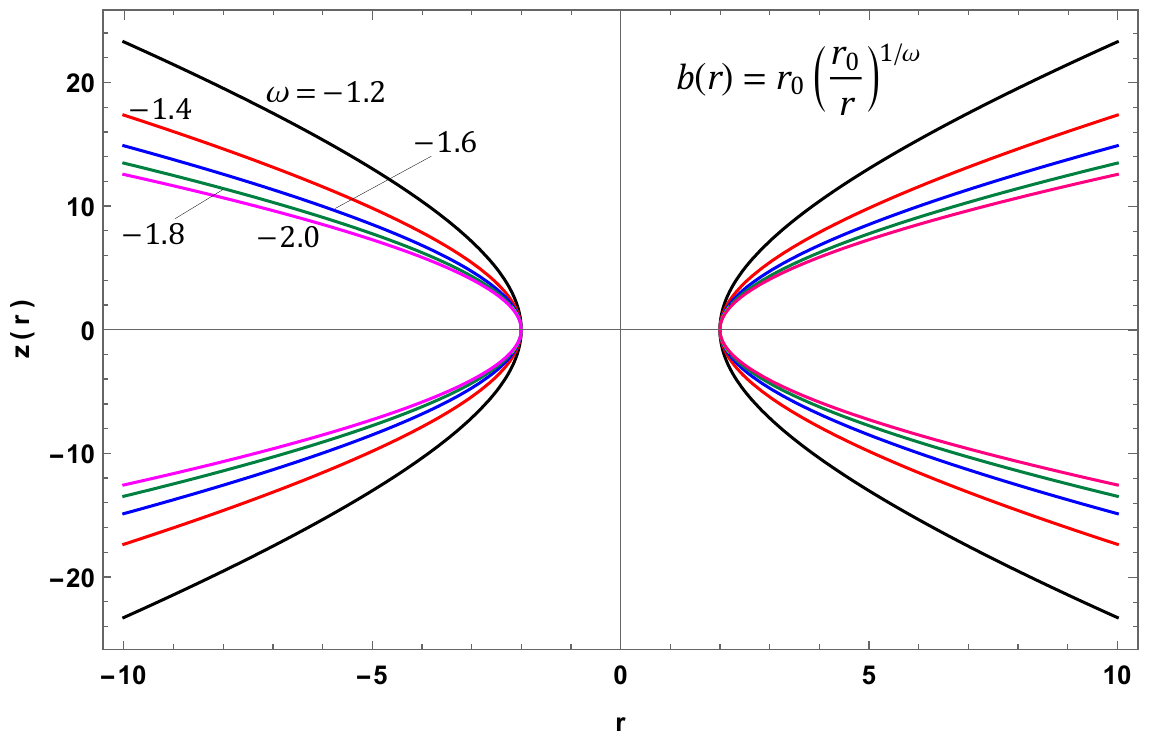}~
\includegraphics[width=11cm,height=7.5cm]{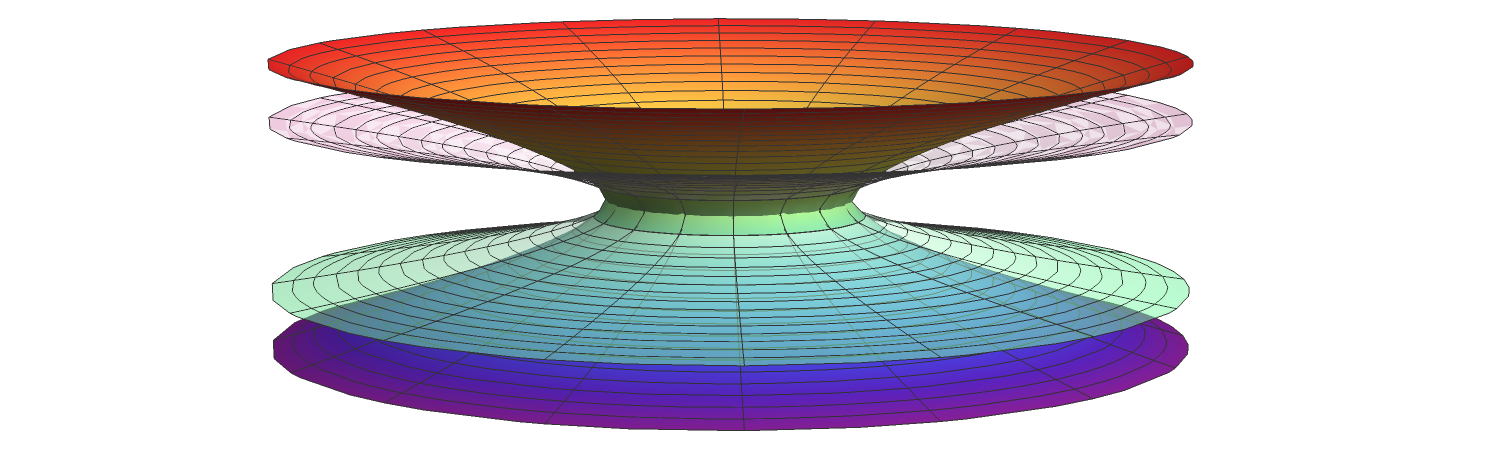}
\caption{Embedding surface for $b(r)=r_{_0}\big(r_{_0}/r\big)^{1/\omega}$ with $r_{_0}=2km$ and its 3D-surface for $\omega=-1.2$ (rainbow) \& $-2$ [WH1].} \label{fig12}
\end{figure*}

\begin{figure*}
\centering
\includegraphics[width=7cm,height=7.3cm]{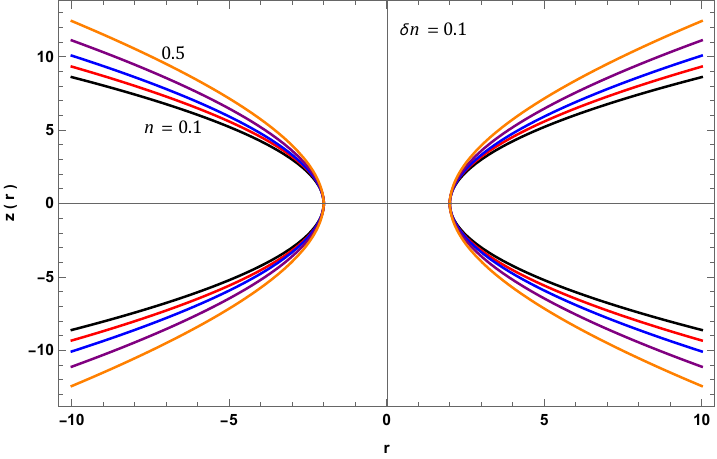}~~~~~
\includegraphics[width=8cm,height=8cm]{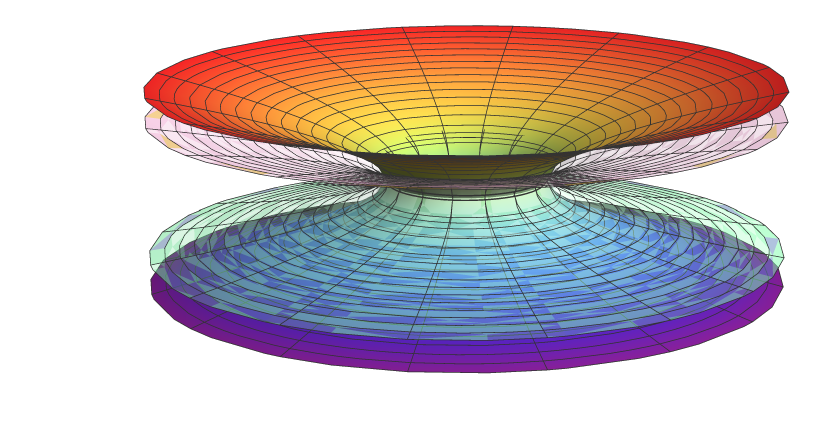}
\caption{Embedding surface for $b(r)=\beta\big(r/\beta\big)^{n}$ with $\beta=2km$ and its 3D-surface for $n=0.1$ \& $0.5$ (rainbow) [WH2].} \label{fig13}
\end{figure*}

\begin{figure*}
\centering
\includegraphics[width=7cm,height=7cm]{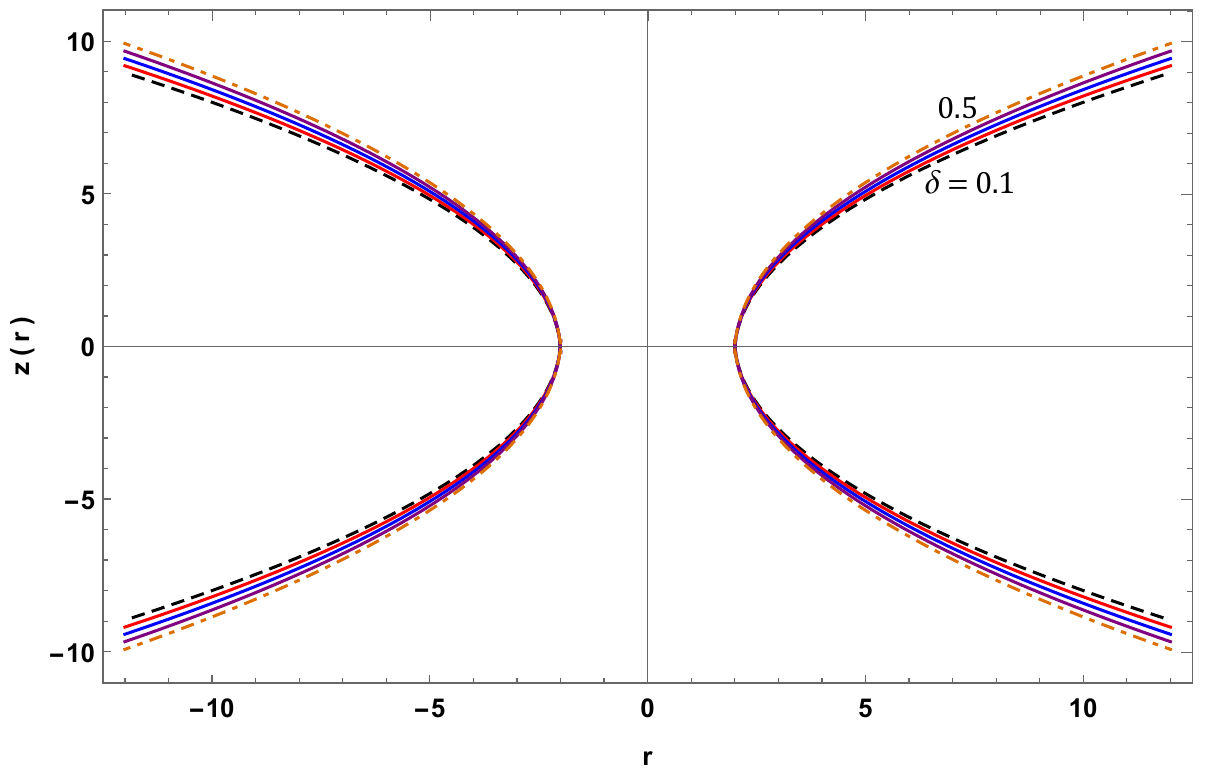}~~~~~~~~
\includegraphics[width=7cm,height=7cm]{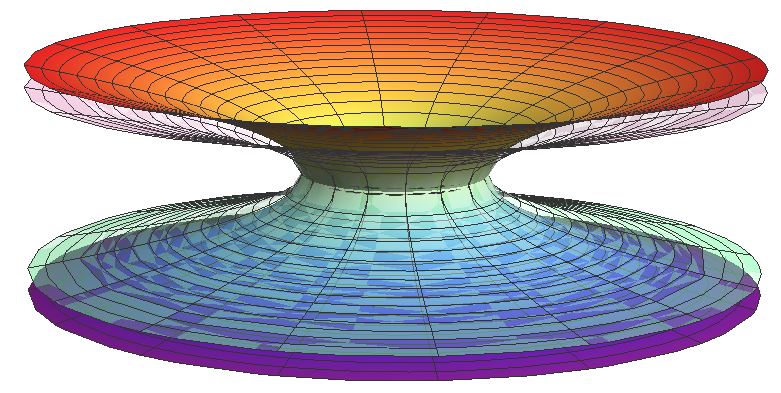}
\caption{Embedding surface for $b(r)=a +\delta  \left(1-\frac{a}{r}\right)$ with $a=2km$ and its 3D-surface for $\delta=0.1$ \& $0.9$ (rainbow) [WH3].} \label{fig14}
\end{figure*}

\begin{figure*}
\centering
\includegraphics[width=7cm,height=7cm]{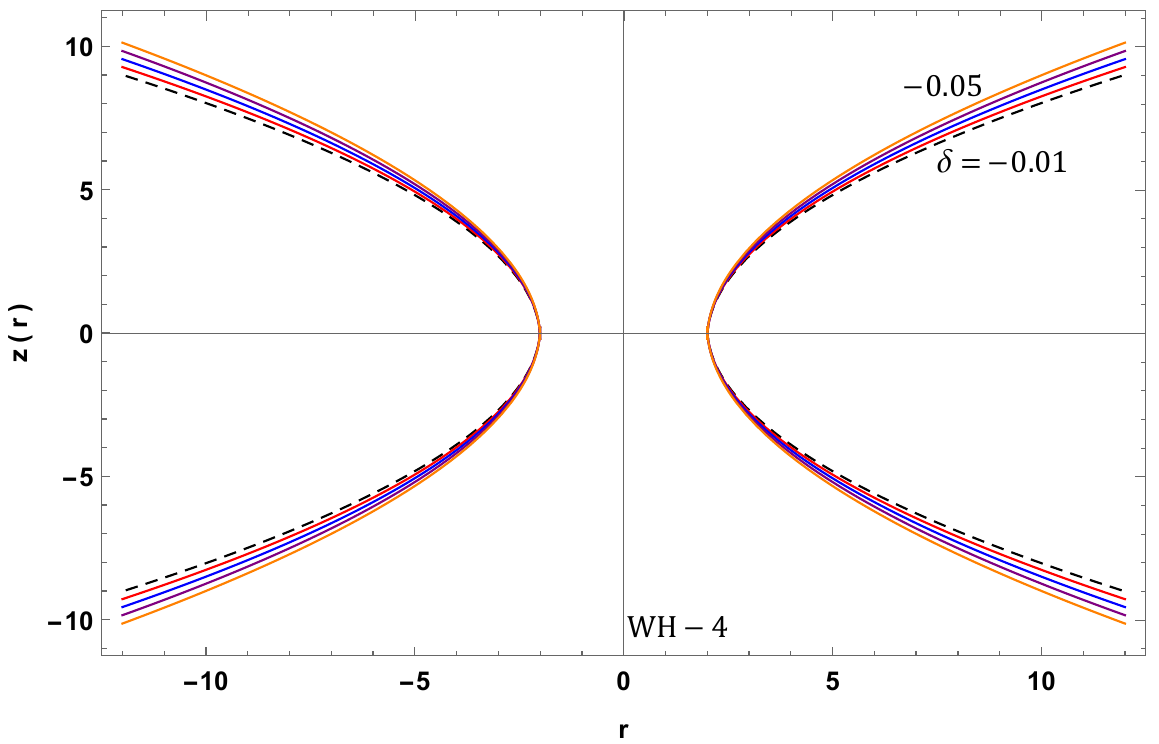}~~~~~~~~~
\includegraphics[width=7cm,height=7cm]{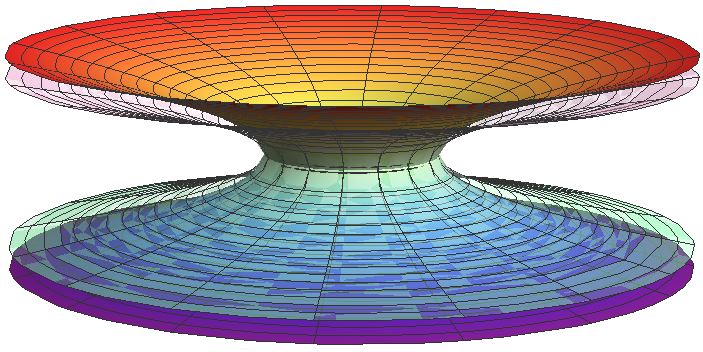}
\caption{Embedding surface for $b(r)=a+a^3\delta  \log \left(\frac{a}{r}\right)$ with $a=2km$ and its 3D-surface for $\delta=-0.01$ \& $-0.09$ (rainbow) [WH4].} \label{fig15}
\end{figure*}

The geometry of the Morris-Thorne WH at equatorial plane ($\phi=\pi/2$) at slice of constant time $t$ is given by
\begin{eqnarray}
ds^{2}=\left[1-\frac{b}{r}\right]^{-1}dr^{2} +r^{2}d\theta^{2} \label{eq53}
\end{eqnarray} 
and embed on 3-dimensional cylinder of the form
\begin{eqnarray}
ds^2 = dr^2+dz^2+r^2 d\theta^2. \label{eq54}
\end{eqnarray}
On comparing \eqref{eq53} and \eqref{eq54} we get
\begin{eqnarray}
1+\left({dz \over dr} \right)^2={1 \over 1-b/r}  ~~~\text{or}~~~\left({dz \over dr} \right)^2={1 \over 1-b/r}-1\nonumber\\={1 \over r/b-1}.~~~~~~
\end{eqnarray}
Now the embedding surface can be determined as
\begin{eqnarray}
z(r) = \pm \int_{r_{_0}}^r {dr \over \sqrt{r/b(r)-1}}.
\end{eqnarray}
This equation is in general not solvable analytically, and therefore numerical methods have to be adopted here. For all the shape functions, the embedding surfaces and it revolutions are shown in Figs. \ref{fig12}, \ref{fig13}, \ref{fig14} and \ref{fig15}. Another important parameter is the proper radial distance defined as
\begin{eqnarray}
\ell(r) = \pm \int_{r_{_0}}^r {dr \over \sqrt{1-b(r)/r}}.
\end{eqnarray}
It is also required to be finite everywhere, and also must be decreasing from the upper universe $\ell \rightarrow +\infty$ to the throat and then smoothly connected with the lower universe $\ell \rightarrow -\infty$ through $\ell=0$. Further, it is also required to fulfill $|\ell(r)|\ge r-r_{_0}$ condition.

\section{Average null energy condition (ANEC)}
\begin{figure*}[!htb]
\centering
\includegraphics[width=7.5cm,height=5.2cm]{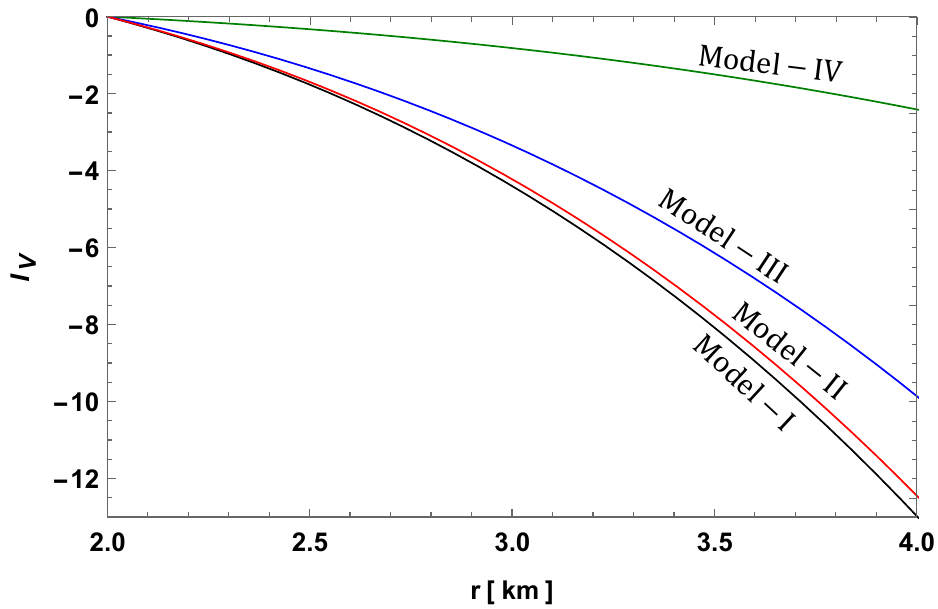}~~~~
\includegraphics[width=7.5cm,height=5.2cm]{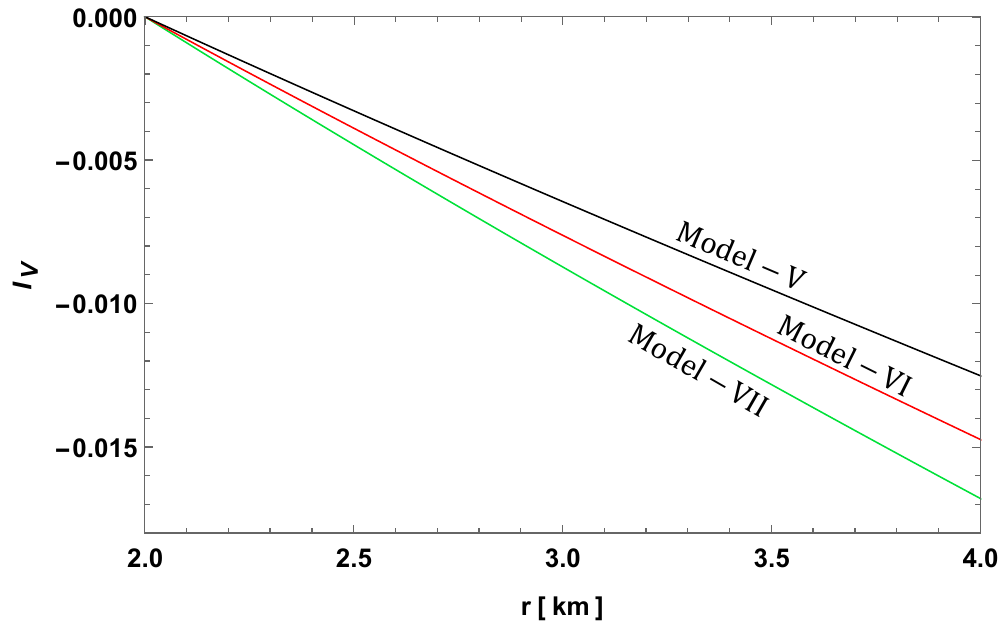}
\caption{Variation of Integral Volume Identifier for the throat radius 2 km for WH models I-VII.} \label{fig16}
\end{figure*}

\begin{figure*}[!htb]
\centering
\includegraphics[width=7.5cm,height=5.5cm]{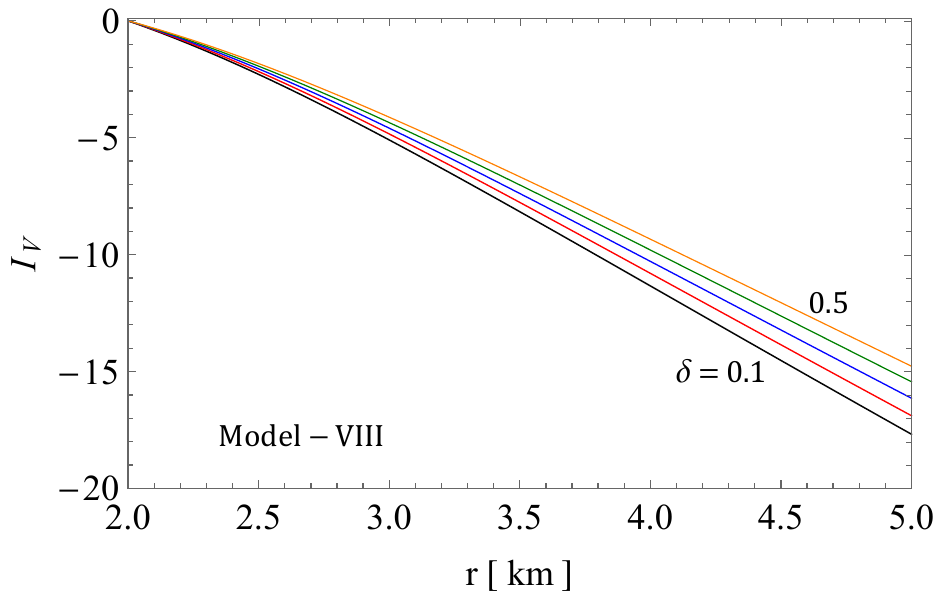}~~~~
\includegraphics[width=7.5cm,height=5.5cm]{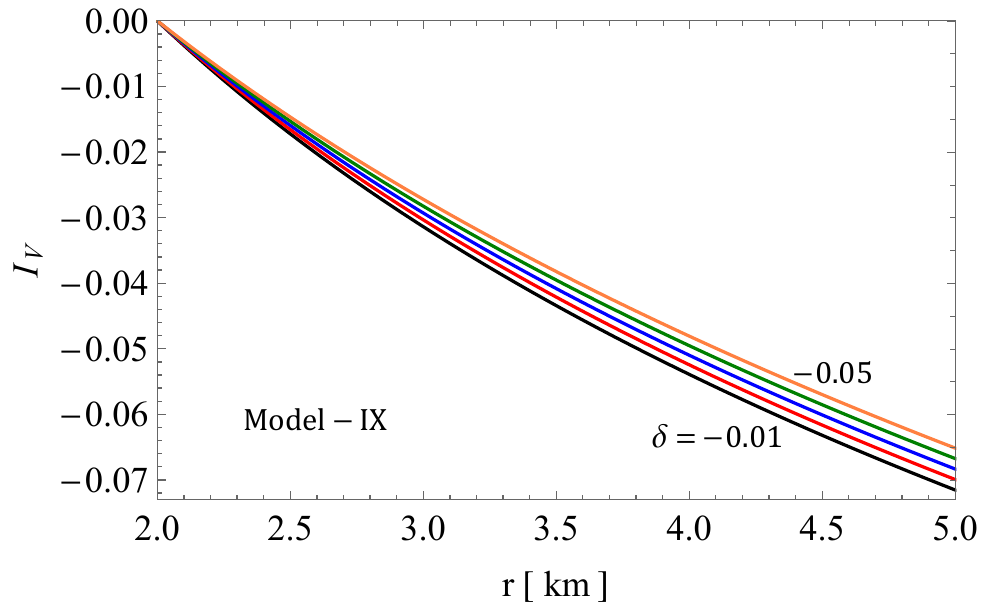}
\caption{Variation of Integral Volume Identifier for the throat radius 2 km for WH models VIII and IX.} \label{fig17}
\end{figure*}

Before finalizing the matter content at the throat of the wormhole is exotic or non-exotic, one must also check for the violation of ANEC defined as
\begin{eqnarray}
\Omega_{ANEC}= \oint T_{\mu \nu} k^\mu k^\nu ~dV ~~~~\forall ~~\text{null vector}~ k^\mu.
\end{eqnarray}
Now, the ANEC along radial null geodesic is given by
\begin{eqnarray}
\Omega_{ANEC}=2\int_{r_{_0}}^\infty \int_0^{\pi} \int_0^{2\pi} (\rho+p)~ r^2 \sin \theta\,dr \,d\theta \,d\phi  \nonumber\\= 8\pi \int_{r_{_0}}^\infty (\rho+p)~ r^2 dr \equiv \mathcal{I}_V.~~~
\end{eqnarray}
This integral is also known as volume integral identifier. In the $\kappa(\mathcal{R},\mathcal{T})-$gravity this integral takes the form
\begin{eqnarray}
\mathcal{I}_V = 8\pi \int_{r_{_0}}^\infty {1 \over \kappa(\mathcal{R},\mathcal{T})} \left[{b' \over r^2}+{2f' \over r}\left\{1-{b \over r}\right\}-{b \over r^3}\right]r^2 dr\nonumber\\~~~~\forall ~~\kappa(\mathcal{R},\mathcal{T})\neq 0.~~~~~~
\end{eqnarray}
which is extremely difficult to solve exactly, since $\mathcal{R}$ and $\mathcal{T}$ comes into play. However, the above expression can be simplified if takes closer to the throat so that $1-b/r<<1$, then
\begin{eqnarray}
\mathcal{I}_V \approx 8\pi \int_{r_{_0}}^{r_{_0}+\epsilon} {1 \over \kappa(\mathcal{R},\mathcal{T})} \left[{b' \over r^2}-{b \over r^3}\right]r^2 dr ~~\forall ~~\kappa(\mathcal{R},\mathcal{T})\neq 0.~~~~~
\end{eqnarray}
which should also be solved numerically. The results of the integration for all the solutions (Models I-IX) can be seen in Figs. \ref{fig16} and \ref{fig17}. From these figures once can conclude that $\mathcal{I}_V \rightarrow 0$ and $r \rightarrow r_{_0}$ for all the cases. This implies that the amount of exotic matter required to open the throats of the wormholes is very small. In spite of the variations of several parameters in all these WH solutions and the profile of the ANEC outside the WH throat are different, its value always vanish at the throat $r=r_{_0}$.

\newpage

\section{Results and Conclusions}
There only exists one article in the literature where WH solutions have been explored in the framework of $\kappa(\mathcal{R},\mathcal{T})$ theory of gravity. Such article was restricted to the case $\kappa(\mathcal{T})=8\pi-\lambda\mathcal{T}$ and did not take into account the non-covariant conservation equation of the stress-energy tensor \eqref{nonconservation3}, which is a necessary condition for physically inspired solutions. The present work is the first article that generates WH solutions via conservation equation. The combination of this conservation equation with field equations can derive WH solutions for three different scenarios: (i) by assuming redshift functions, (ii) by assuming an equation of state and (iii) by assuming shaped functions. The first solution was explored by assuming a constant redshift function ($f'(r)=0$) that generates a shape-function $b(r)=-Cr^3$, which is asymptotically non-flat i.e. $b/r \nrightarrow 0$ at $r \rightarrow \infty$. Hence, not a physical solution, however it can become a physical one by matching it with an exterior solution e.g. Schwarzschild vacuum. Further, other solutions may also be generated by assuming other forms of redshift functions, which is outside the current article.

In the second approach, we have considered a linear equation of state $p=\omega \rho$ in the conservation equation and able to generate a shape-function $b(r)=r_{_0}(r_{_0}/r)^{1/\omega}$. It is physically inspired WH solution for all $\omega < -1$ and also asymptotically flat as well. Using this shape-function we have solved for redshift function through conservation equation. The nature of the pressure, density and null energy condition
under three forms of $\kappa(\mathcal{R},\mathcal{T})$ as $\kappa(\mathcal{T})$, $\kappa(\mathcal{R})$ and $\kappa(\mathcal{R},\mathcal{T})$ through coupling constants were provided.

Next, we have examined few other forms of WH solutions under $\kappa(\mathcal{T})$, $\kappa(\mathcal{R})$ and $\kappa(\mathcal{R},\mathcal{T})$ gravity. Model I WH has constant density and pressure along with the violation of NEC. Model II has decreasing density and pressure, while Model III is slightly increasing in density near the throat and flat when far away from $r=r_{_0}$, while its pressure is negative and constant. The Model IV has increasing density and pressure outward, which might be due to direct geometric-matter coupling non-linearly through $\kappa(\mathcal{R},\mathcal{T})$. Surprisingly, Model V, VI and VII have exactly same density and pressure which are decreasing outward. Similarly, Model VIII and IX also have decreasing density and pressure. All these WHs (Model II-IX) satisfy the flaring-out condition, asymptotically flat and violates NEC. Since their equation of state parameters $\omega$ are less than -1, the throat is supported by ``phantom energy'' that provides a negative pressure.

The equatorial view of the WHs for a constant time slice are discussed in the next section that helps to the generate the embedding surface and its revolution (Figs. \ref{fig12}-\ref{fig15}). For the WH1, the throat length decreases when $\omega$ decreases from $-1.2$ to $-2.0$. On the other hand, the throat length increases when $n$ goes from 0.1 to 0.5 in WH2. In the same trend, WH3 throat length also increases with the increase in $\delta$ from 0.1 to 0.5. The last model WH4, throat length increases with decrease in $\delta$ from $-0.01$ to $-0.09$. As long as the parameter $\omega<-1$, all the wormhole shape functions satisfy all the required physical conditions. However if $\omega > -1$, the flaring-out condition begins being violated and therefore the solutions cannot represent physical WH solutions. For WHII, the density and pressure has no significant effect when the $\beta$-parameter changes. As $\lambda$ increases, the density increases and the pressure becomes repulsive in WH1, while both remain unchanged in WH-V. In the case of WH2, as $\beta -$parameter increases the density decreases and the pressure increases. For WH2, the density and pressure has no significant effect when the $\beta -$parameter changes. For WH4 and WH7, the density and pressure decrease with increasing $\gamma-$parameter, while the wormhole remains unaffected.

Since all these WH models are supported by phantom energy ($\omega<-1$), one can estimate how much phantom energy is required to support the throat of these WHs. This can be estimated by the calculating the ANEC via volume integral identifier $\mathcal{I}_V$ at the very throat of the wormholes. It is very clear seen from Figs. \ref{fig16} and \ref{fig17} that ANEC vanishes exactly at $r_{_0}$ for all these WH models. This implies that ANEC holds at the throat of these WHs and therefore minimum exotic matter is required to hold it. Finally, it is concluded that the WH models presented here hold conservation equation, asymptotically flat (except constant redshift function model), holds flaring-out condition, supported by phantom energy and ANEC holds at the throat.

\end{document}